\documentclass[preprint]{aastex}
\usepackage{mkfig}

\begin{document}

\title{\bf Coronal Emission Measures and Abundances for Moderately
  Active K Dwarfs Observed by {\em Chandra}}

\author{Brian E. Wood, Jeffrey L. Linsky}

\affil{JILA, University of Colorado, Boulder, CO
  80309-0440; woodb@origins.colorado.edu, jlinsky@jila.colorado.edu.}

\begin{abstract}

     We have used {\em Chandra} to resolve the nearby 70~Oph (K0~V+K5~V)
and 36~Oph (K1~V+K1~V) binary systems for the first time in X-rays.
The LETG/HRC-S spectra of all four of these stars are presented and
compared with an archival LETG spectrum of another moderately active
K dwarf, $\epsilon$~Eri.  Coronal densities are estimated from O~VII line
ratios and emission measure distributions are computed for all five of
these stars.  We see no substantial differences in coronal density
or temperature among these stars, which is not surprising considering
that they are all early K dwarfs with similar activity levels.
However, we do see significant differences in coronal abundance patterns.
Coronal abundance anomalies are generally associated with the
first ionization potential (FIP) of the elements.  On the Sun,
low-FIP elements are enhanced in the corona relative to high-FIP
elements, the so-called ``FIP effect.''  Different levels of FIP effect
are seen for our stellar sample, ranging from 70~Oph~A, which shows a
prominent solar-like FIP effect, to 70~Oph~B, which has no FIP bias at
all or possibly even a weak inverse FIP effect.
The strong abundance difference exhibited by
the two 70~Oph stars is unexpected considering how similar these stars
are in all other respects (spectral type, age, rotation period, X-ray
flux).  It will be difficult for any theoretical explanation for the FIP
effect to explain how two stars so similar in all other respects can
have coronae with different degrees of FIP bias.
Finally, for the stars in our
sample exhibiting a FIP effect, a curious difference from the solar
version of the phenomenon is that the data seem to be more consistent
with the high-FIP elements being depleted in the corona rather than a
with a low-FIP enhancement.

\end{abstract}

\keywords{stars: abundances --- stars: coronae --- stars:
  late-type --- X-rays: stars}

\section{INTRODUCTION}

     Extensive surveys by the {\em Einstein} and {\em ROSAT} X-ray
satellites have shown that all late-type main sequence stars are
surrounded by $T\geq 10^{6}$~K coronae analogous to the more easily
observed corona surrounding our own Sun \citep[e.g.,][]{js04}.
Although the mechanism or mechanisms responsible for heating these
outermost stellar atmospheric regions to such high temperatures are not
entirely understood, it is clear that coronal heating and coronal structure
are controlled by magnetic activity.  On the Sun, magnetic fields are
associated with a number of fascinating phenomena:  sunspots, flares,
the solar wind, coronal mass ejections, prominences, etc.  Lack of spatial
resolution makes most of these phenomena difficult to study on stars, so
most of what we know about stellar magnetic activity comes solely by
studying their global radiation.  Since X-rays are produced by
stellar coronae, stellar X-ray luminosities have become one of the
primary measures of coronal activity, and much work has been done to see
how activity as measured by X-ray emission correlates with other stellar
properties such as age and rotation.  A thorough review of X-ray studies
of stellar coronae is provided by \citet{mg04}.

     X-ray spectroscopy provides more diagnostic information on the
characteristics of stellar coronae than the X-ray luminosity.
The numerous pulse height spectra obtained by satellites such as
{\em Einstein}, {\em ROSAT}, and {\em ASCA} provide some information
about the X-ray spectra of many stars, but their very low spectral
resolution ($R\equiv \lambda/\Delta\lambda \lesssim 50$) severely limits
the usefulness of such data.  In contrast, the grating spectra
produced by the now defunct {\em Extreme Ultraviolet Explorer} (EUVE),
and by the current {\em Chandra} and {\em XMM-Newton} missions
can resolve individual lines, providing a more
complete arsenal of spectroscopic techniques for inferring coronal
characteristics.

     For example, such spectra allow more precise determinations of
coronal emission measure distributions, which indicate the temperature
distribution of the emitting plasma.  These spectra often contain
density-sensitive lines that allow the measurement of coronal densities
\citep[e.g.,][]{jun02}.  Finally, elemental abundances in stellar
coronae can be estimated from X-ray spectra.  Much attention has been
focused on these abundances, since coronae often exhibit curious abundance
patterns that depend on first ionization potential (FIP).  For the Sun and
other stars of modest or low activity, elements with low FIP are often
observed to have enhanced abundances relative to high FIP elements
\citep{uf00}.  This is the so-called ``FIP effect.''  However,
coronae of very active stars are often found to behave differently, either
having no FIP effect or even an inverse FIP effect where low FIP elements
are depleted rather than enhanced in the corona relative to high FIP
elements \citep[e.g.,][]{ma03}.  The origin of these coronal
abundance effects is an active area of theoretical study
\citep[e.g.,][]{jml04}.

     We report here on X-ray spectra obtained by {\em Chandra}'s Low
Energy Transmission Grating (LETG), using the HRC-S detector.  Our
new observations are of two moderately active K dwarf binaries,
70~Oph (K0~V+K5~V) and 36~Oph (K1~V+K1~V).  {\em Chandra}'s superb
spatial resolution allows these binaries to be resolved for the first
time in X-rays, meaning that these two observations provide spectra of
all four stars in the two binary systems.  In addition to these data,
we also analyze archival LETG/HRC-S spectra of another moderately active
K dwarf, $\epsilon$~Eri (K1~V).  The properties of the five stars in our
sample are summarized in Table~1.

     Our goal is to compare the X-ray spectra of these five very similar K
dwarfs with very similar coronal activity levels, in order to see whether
these stars have identical coronal properties.  In the case of the 70~Oph
and 36~Oph binaries, we can actually compare spectra of stars that are
coevol, and therefore have identical ages, presumably identical
photospheric abundances, and rotation rates that are observed to be
nearly identical in both cases (see Table~1).  Another reason behind
our choice of targets is that these are among the few solar-like
stars with measured mass loss rates, which provides us the opportunity to
compare coronal properties with the strengths of the stellar winds that
emanate from these coronae.

       The wind measurements, which are based on H~I
Lyman-$\alpha$ absorption from the interaction regions between the winds
and the interstellar medium \citep{bew05a}, suggest that despite
their similarities the three stellar systems have significantly different
mass loss rates.  With respect to the solar mass loss rate of
$\dot{M}_{\odot}=2\times 10^{-14}$ M$_{\odot}$~yr$^{-1}$, the stellar
mass loss rates range from $15~\dot{M}_{\odot}$ for 36~Oph up to
$100~\dot{M}_{\odot}$ for 70~Oph (see Table~1).  (Unfortunately, the
wind-ISM interaction regions that are used to measure the stellar winds
encompass both stars in the 36~Oph and 70~Oph systems, so wind
measurements can only be made for the combined winds of these
binaries.)  We will see whether there are differences in the coronal X-ray
spectra of these stars that may be connected with these differences in
wind strength.

\section{OBSERVATIONS AND DATA ANALYSIS}

     The {\em Chandra} LETG/HRC-S observations are summarized in
Table~2.  Both 36~Oph and 70~Oph are binary systems with
separations of $\sim 5^{\prime\prime}$.  {\em Chandra} easily resolves
these binaries, but the standard pipeline processing does not properly
separate the spectra of the two stars.  Thus, we have processed the data
ourselves using version 3.2.1 of the CIAO software, and version 3.0.1 of
the CALDB calibration database.  This reprocessing was not strictly
necessary for $\epsilon$~Eri, but we did it anyway for the sake of
consistency.  We followed appropriate analysis threads provided by the
{\em Chandra} Science Center (see
http://asc.harvard.edu/ciao/threads/gspec.html).  Although the HRC-S
detector has little intrinsic energy resolution, the
pulse height information that is provided can still be used to reject
a non-trivial number of background counts, thereby reducing detector
background.  Thus, the CIAO data reduction includes this correction using
the recommended ``light'' background filter.

     Observations with the LETG/HRC-S combination yield a zeroth order
image of the target at the aimpoint, with two essentially identical X-ray
spectra dispersed in opposite directions along the long axis of the HRC-S
detector (i.e., a plus order and a minus order spectrum).  The spectrum
is dominated by the first order, but weaker higher orders
are superimposed on the first order spectrum.  The 36~Oph and 70~Oph
observations were each scheduled when the binary axis was
roughly perpendicular to the dispersion direction, in order to produce
two parallel stellar spectra that are separated as much as possible.
The spectra cover an effective wavelength range from $5-175$~\AA, with a
resolution of $FWHM\approx 0.06$~\AA\ throughout.

     The spatial resolution in the cross-dispersion direction worsens
at long wavelengths where the spectrum is furthest from the focal point.
For this reason the standard default spectral extraction window has a
bow tie shape that increases in width at the long wavelength ends of both
plus and minus spectral orders, with a similarly shaped background
region surrounding it.  However, using such an extraction window for the
36~Oph and 70~Oph data would lead to source confusion due to the presence
of more than one source.  Thus, we experiment with different
extraction windows.  We ultimately decide to use a conservatively broad
window with a width of 30 pixels centered on each stellar spectrum for
$\lambda <90$~\AA.  It is necessary to broaden the window for
$\lambda >90$~\AA.  To avoid overlapping source windows for the two
components of our binaries, for each star we expand the window only in the
direction away from its companion star by 30 pixels.  At the longest
wavelengths, we have to accept that there is some degree of unresolvable
blending.  However, there are very few lines detected with
$\lambda >90$~\AA\ for 36~Oph and 70~Oph (see \S4), so this is tolerable.
Since $\epsilon$~Eri is not a binary, we can simply expand the extraction
window for $\lambda >90$~\AA\ in both directions to a full width of 90
pixels.  In addition to the source windows, we extract spectra from two
120-pixel windows on each side of the stellar spectra to serve as
background measurements.

     After processing the data with the CIAO software, we coadd the
resulting plus and minus order spectra.  For each star, a wavelength shift
of 1 or 2 pixels (1 pixel=0.0125~\AA) has to be applied to either the plus
or minus order to ensure that the two orders are properly aligned
before coaddition.  The detector background is then subtracted from the
data using the background spectra described above.

\section{THE ZEROTH ORDER IMAGES}

     Before showing the spectra, it is useful to consider first the
zeroth order images that are also provided by the LETG/HRC-S observations.
Figure~1 shows these images for the 36~Oph and 70~Oph binaries.
{\em Chandra} is the first X-ray satellite capable of truly resolving these
star systems.  Prior to {\em Chandra}, the best spatial resolution
available was from {\em ROSAT}'s High Resolution Imager (HRI).
ROSAT/HRI observed both 36~Oph and 70~Oph \citep{js04}.  In
each case, an asymmetry in the HRI image has the appropriate orientation
to suggest that both stars are contributing to the emission, but
it is impossible to say more than that from those data.  In contrast,
Figure~1 shows how easily {\em Chandra} resolves the binaries.
For 36~Oph, we measure a separation of $5.132\pm 0.007^{\prime\prime}$
and a position angle of $141.01\pm 0.12^{\circ}$, while for 70~Oph these
quantities are $4.708\pm 0.005^{\prime\prime}$ and $137.75\pm 0.09^{\circ}$,
respectively.  The stellar separation shown in Figure~1 is also
indicative of the separation of the spectra at short wavelengths, though
as mentioned in \S2 the spatial resolution deteriorates at long
wavelengths.  For both binaries, the A component is slightly brighter,
with 36~Oph~A and 70~Oph~A accounting for 58.2\% and 60.4\% of the
total X-ray flux, respectively.

     The zeroth order images also provide light curves for our targets,
which are useful to see whether our spectra are affected by any flaring.
The light curves are shown in Figure~2.  Some modest,
gradual variability with timescales of hours is present for our targets,
but the closest thing to a flare is the peak seen for 36~Oph~B at
$t=31$~ksec, where the flux briefly increases by about a factor of 2.
Even if this is a weak flare, it is not strong enough to significantly
affect the nature of the spectrum, so no attempt is made to remove it
from the data.

\section{SPECTRAL ANALYSIS}

\subsection{Line Identification and Measurement}

     The final processed spectra of our five K dwarf stars are shown in
Figure~3.  The spectra are rebinned by a factor of 3 to improve
signal-to-noise (S/N).  We use version 4.2 of the CHIANTI atomic database
to identify emission lines \citep{kpd97,pry03}.  The
identified lines are shown in Figure~3 and listed in Table~3.  Table~3
lists counts for the detected lines, which are measured by direct
integration from the spectra.  Line formation temperatures are quoted in
the third column of the table, which are based on maxima of contribution
functions computed using the ionization equilibrium computations of
\citet{ma85}.  Uncertainties quoted in Table~3 are
1$\sigma$.  For all undetected lines, we quote 2$\sigma$ upper limits
based on summing in quadrature the uncertainties in 5 bins surrounding
the location of the line, and then multiplying this sum by 2.

     We measure all lines that appear visually to be at least
marginally detected.  However, all cases in which the uncertainty
implies a $<2\sigma$ detection must be considered questionable.
After the analysis described in \S4.4 is completed, we confirm
{\em a posteriori} that these questionable detections are at least
plausible based on the derived emission measure
distributions and abundances (which are constrained primarily by
the stronger emission lines in the analysis).  Many of the emission
lines are blends.  We list in Table~3 all lines that we believe contribute
a significant (i.e., $\gtrsim 10\%$) amount of flux to the feature based
on the line strengths in the CHIANTI database.  This determination is
reassessed after emission measure distributions are estimated, and model
spectra can be computed from them and compared with the data (see \S4.4
and Fig.~3).  A few of the features identified in Figure~3 are blends of
lines of different species (see Table~3).  Although we measure counts for
these blends and list them in Table~3, these measurements are not used in
any of the following analysis.

\subsection{Coronal Densities}

     One coronal property that can in principle be inferred from
high-resolution X-ray spectra is the density.  The best density
diagnostics available in our spectra are various He-like triplets:
Si~XIII $\lambda$6.7, Mg~XI $\lambda$9.2, Ne~IX $\lambda$13.6,
and O~VII $\lambda$21.8.  In order of increasing wavelength, these
triplets each consist of a resonance line, an intersystem line, and a
forbidden line.  The ratio of the intersystem and forbidden lines
is density sensitive.

     Unfortunately, the S/N of the Si~XIII, Mg~XI, and Ne~IX triplets
in our spectra is generally not sufficient to allow
for a precise density measurement.  Due to blending, we simply list
in Table~3 the total flux from the Si~XIII and Mg~XI triplets, and for
Ne~IX there is no attempt to separate the fluxes of the resonance and
intersystem lines.  For $\epsilon$~Eri, there {\em is} sufficient
S/N to separate these lines using profile-fitting techniques and
careful correction for blends (e.g., numerous Fe~XIX
blends for the Ne~IX triplet).  \citet{jun02} have already
performed this analysis for $\epsilon$~Eri, so we do not repeat it
here.

     Since the O~VII triplet is the only case where the lines are
sufficiently separated and observed with reasonable S/N for all of our
stars, we focus our attention on these lines.  Line ratios of
$f/i\equiv \lambda22.101/\lambda21.807$ are computed from the
fluxes in Table~3, and we then use the same prescription
as \citet{jun02} to convert these ratios to electron densities
(in cm$^{-3}$), which utilizes the results of collisional equilibrium
models from \citet{dp01}.  The resulting densities are listed in
Table~4.  For 36~Oph~B and the two 70~Oph stars,
the line ratios are consistent with the low density limit of the
diagnostic, so we can only quote upper limits for $n_{e}$.
We note that our $\epsilon$~Eri measurement is consistent with the
$\log n_{e}=10.03\pm 0.23$ result from \citet{jun02} based on
these same O~VII lines.  A solar coronal density derived from these
same O~VII lines, $\log n_{e}=9.4$, is also listed in Table~4 for
the sake of comparison.  This value is based on the average $f/i$
ratios observed from a collection of rocket measurements of various
active regions \citep{dlm87}.  With the possible exception of
36~Oph~A, the coronal densities of our moderately active K dwarfs
are not clearly different from the solar value.

     All five spectra are basically consistent with a coronal density of
$\log n_{e}\approx 10.0$.  There is some suggestion that 36~Oph~A
might have a slightly higher density.  The smaller $f/i$ ratio
for 36~Oph~A is apparent in Figure~3.  However, the uncertainty in this
ratio is too large to clearly demonstrate higher densities, and the
error bar of the $\log n_{e}=10.55\pm 0.51$ measurement for
36~Oph~A overlaps the values or upper limits quoted for the other stars.
Thus, we conclude that there is no convincing evidence
that the coronal densities of our five stars are different.

\subsection{Flux Ratios}

     In \S4.4, we compute absolute abundances and emission measure
distributions.  However, since our primary science goals are comparative
in nature (see \S1), we can learn a lot by simply comparing the
observed line fluxes of our stars with each other without going through
the more complex analysis of \S4.4.  Thus, in Figure~4 we compare the
line fluxes of various pairs of stars by plotting line flux ratios
versus line formation temperatures, using different symbols and colors to
indicate high-FIP (${\rm FIP}>10$ eV) and low-FIP (${\rm FIP}<10$ eV)
elements.  In cases where there is more
than one line of a given species, we simply add fluxes of all lines
detected for both stars before computing the flux ratio.

     In order to expand the temperature regime explored in this
analysis, we have supplemented our {\em Chandra} line measurements
with UV measurements from spectra taken by the {\em Hubble Space
Telescope} (HST).  These observations, which utilize the Space Telescope
Imaging Spectrometer (STIS) instrument on HST, are available for
$\epsilon$~Eri, 70~Oph~A, and 36~Oph~A.  The STIS observations of
$\epsilon$~Eri and 70~Oph~A use the E140M grating, covering a
wavelength range of $1150-1730$~\AA\ at a spectral resolution of
$R\approx 45,000$, while the available
36~Oph~A spectrum uses the E140H grating, covering a wavelength range of
$1160-1357$~\AA\ at a higher resolution of $R\approx 110,000$.
The UV line measurements are listed in Table~5 along with line formation
temperatures from \citet{ma85}.

     All but one of the HST lines are formed either in the chromosphere
($3.8<\log T<4.3$) or transition region ($4.3<\log T<5.8$), but there is
one coronal line (with $\log T>5.8$), the Fe~XII line at 1242.01~\AA.
The $\epsilon$~Eri and 70~Oph~A fluxes for this line are taken directly
from \citet{tra03}.  Most of the fluxes in Table~5 are our
own measurements of the archival HST spectra based on direct flux
integration of the lines, with the following exceptions.  Most
of the $\epsilon$~Eri fluxes are taken from \citet{sas05},
and the fluxes of the H~I Lyman-$\alpha$ line at 1215.671~\AA\ are taken
from \citet{bew05b}, who have corrected for interstellar
absorption.

     The flux ratio plots in Figure~4 illustrate many interesting
similarities and differences in the spectra of our K dwarfs.
Panel (c) compares fluxes of the two 36~Oph stars.  The flux ratios
are generally consistent with the flux ratio seen in the zeroth-order
{\em Chandra} image of the binary (see \S3), with one notable exception.
For the O~VII lines, the 36~Oph~A/36~Oph~B ratio is lower than one would
expect.  This is also suggested by the noticeably different
O~VIII/O~VII ratios for these 2 stars apparent in Figure~3.  This may
be indicative of small differences in the coronal emission measure
distributions, with 36~Oph~A having a slightly hotter corona.

     Much more dramatic differences are seen in Figure~4a, which
compares the line fluxes of the two 70~Oph stars.  The 70~Oph~A/70~Oph~B
ratios are higher for the low-FIP lines than the high-FIP lines,
the implication being that 70~Oph~A's corona apparently has a much
more prominent FIP effect than its companion star.  The spectral
differences between the two stars that lead to this result are readily
apparent in Figure~3.  (Compare, for example, the relative strengths of
the Fe~XVII and O~VIII lines.)  This substantial
difference in coronal FIP effect is very surprising considering just
how similar the two 70~Oph stars are.  Since they are members of the same
star system, they clearly have the same age, and these stars
also have very similar spectral types, rotation rates, and X-ray
surface fluxes (see Table~1).  Why do these coronae then
exhibit different levels of FIP bias?  We will return to this issue
in \S5.1.

     Figure~4b indicates that 70~Oph~A not only has a stronger FIP
bias than 70~Oph~B, but also clearly has a stronger FIP effect than
$\epsilon$~Eri, at least in the corona.  A relative lack
of available low-FIP lines at lower temperatures means that it is not
entirely clear whether this difference also exists at transition
region temperatures, but the effect certainly seems
to be stronger in the corona.  The origin of the FIP effect is generally
assumed to lie in the chromosphere, where the low-FIP elements are
ionized but the high-FIP elements are not.  Thus, one would naively
expect the FIP effect to be just as apparent in the transition region
as it is in the corona.  But Figure~4b suggests that the relative
strength of a FIP effect can be different between the corona and
transition region.  This is consistent with solar observations,
which show a curious absence of a FIP effect in the transition region
\citep{jml95}.  Perhaps this difference is
symptomatic of the transition region emission arising in large part from
different magnetic structures than the coronal emission, a conclusion
supported by images of the Sun's corona and transition region, which
have very different appearances \citep{uf94}.

     For both 70~Oph~A and 36~Oph~A, line ratios with $\epsilon$~Eri are
generally lower in the corona than in the transition region (see
Figs.~4b and 4d).  This is consistent with a higher degree of coronal
heating for $\epsilon$~Eri, which is perhaps not surprising since
$\epsilon$~Eri is a faster rotator and is slightly more active than
70~Oph~A and 36~Oph~A (see Table~1).

\subsection{Emission Measure and Abundance Analysis}

     For a more quantitative assessment of the abundances and temperatures
of the coronal plasma responsible for the X-ray emission, we perform an
emission measure analysis for our five stars based on the line measurements
listed in Table~3.  This analysis requires assumptions of collisional
ionization equilibrium, Maxwellian velocity distributions, and uniform
abundances throughout the corona.  The emission measure distribution
(in units of cm$^{-3}$) is defined as
\begin{equation}
EM(T)\equiv n_{e}^{2} \frac{dV}{d\log T},
\end{equation}
where dV is a coronal volume element.  The observed flux for a
given line can then be expressed as
\begin{equation}
f=\frac{1}{4\pi d^{2}} \int G(T,n_{e})EM(T) d\log T,
\end{equation}
where $d$ is the stellar distance and $G(T,n_{e})$ is the line contribution
function, which includes both the line emissivity and the assumed
elemental abundance of the atomic species in question.

     Computing a line flux from a known emission measure distribution and
with a known elemental abundance is trivial using equation (2).  However,
the inverse problem of inferring EM(T) and elemental abundances from a set
of observed line fluxes in a self-consistent manner is challenging, and
a single, unique solution is not strictly obtainable.
We use version 2.1 of the PINTofALE software developed by
\citet{vk00} to perform these calculations, which tackles the
inverse problem using a Markov Chain Monte Carlo approach \citep{vk98}.

     Before using the measured line counts in Table~3 in the emission
measure analysis, we must divide them by the exposure times in Table~2,
and we must also correct for interstellar absorption.  Interstellar H~I
column densities are listed in Table~1 for all of our targets.  These
columns are measured from H~I Lyman-$\alpha$ absorption \citep{bew05b}.
The column densities are converted to wavelength dependent
transmittance curves using the prescription of \citet{rm83},
which provide the necessary information to correct line
fluxes for ISM absorption.

     Finally, wavelength-dependent effective area curves must be derived
for our spectra.  The spacecraft dithering that is done during
every {\em Chandra} observation to smooth detector nonuniformities
results in slightly different effective areas from one observation to
the next; the exact dithering pattern yields a different aspect
solution for converting detector coordinates to spatial coordinates.
Deviance from the standard effective area curve is especially noticeable
near the two gaps in the HRC-S detector (at $\sim 53$~\AA\ for the minus
order spectrum and $\sim 63$~\AA\ for the plus order spectrum), because
their exact wavelengths depend on the source position as well as on the
aspect solution.  Thus, effective area curves must be computed separately
for each of our sources, which we do using the appropriate CIAO routines.
Since the lines listed in Table~3 are all first order lines, we only
have to compute first order effective area curves for the emission
measure analysis.  However, we also compute effective areas for
orders 2--5 to assist in the computation of the synthetic spectra
shown in Figure~3 (see below).

     All lines listed in Table~3 are considered in the emission measure
analysis, with a few exceptions.  Blends of lines of different species
are not considered.  In cases of blends of lines of the same species, we
use the predicted line strengths from CHIANTI to divide the flux among
the blended lines.  The density sensitive lines of the He-like triplets
are not included (see \S4.2).  The PINTofALE software allows for
consideration of upper limits, so the 2$\sigma$ upper limits listed
in Table~3 {\em are} taken into account in the analysis.  To improve
constraints on the emission measure distributions, we also include the
Fe~XII $\lambda$1242 line from the HST spectra (see Table~5).  The HST
and {\em Chandra} data are not simultaneous, so variability is potentially
a problem for the Fe~XII measurements.  However, the Fe~XII data points in
Figures~4b and 4d are basically consistent with the {\em Chandra} data
points, so we do not believe this is a major issue.

     The CHIANTI database is the source of our line emissivities
\citep{kpd97,pry03}.  We assume a density of
$\log n_{e}=10$, which is consistent with the measurements in \S4.2 for
all of our stars.  Temperature-dependent ionization fractions are taken
from \citet{pm98} based on their collisional equilibrium
calculations.  The solar photospheric abundances of \citet{ng98}
are used for our initial assumed abundances.  For elements with
detected lines we can compute abundances during
the course of the emission measure analysis.  However, line
measurements alone only allow {\em relative} abundances to be computed,
meaning the derived emission measure distribution cannot be
normalized to an absolute value (though the shape of the distribution
can be established).  Thus, in the initial emission measure computation
the Fe abundance is fixed at the solar photospheric value, and other
abundances are measured relative to that.  In order to determine the
{\em absolute} Fe abundance and thereby properly normalize the
emission measure distribution, the line-to-continuum ratio must be
assessed, a process that will be described below.

     Figure~5 shows the emission measure distributions derived by the
PINTofALE software using 0.1 dex increments in $\log T$.  The derived
abundances relative to Fe are listed in Table~4.  The emission measure
uncertainties shown in Figure~5 and the abundance uncertainties quoted
in Table~4 are 90\% confidence intervals.

     As mentioned above, the
only way to determine absolute Fe abundances, Fe/H, and thereby properly
normalize the emission measure distributions is to assess the relative
strengths of the line and continuum emission.  Figure~6 shows how this
is done, using the $\epsilon$~Eri spectrum as an example.  The bottom
panel shows a smoothed version of the $\epsilon$~Eri spectrum, and
several model spectra computed assuming different values for Fe/H.
Higher Fe/H values correspond to lower emission measures, so higher
Fe/H leads to a lower continuum.  In deciding what Fe/H value leads
to the best fit, we believe it is better for the model spectrum to
underestimate the observed fluxes than to overestimate it, since missing
lines in the atomic database could potentially explain an underestimate.
With this in mind, we decide that ${\rm [Fe/H]}=0.7{\rm [Fe/H]}_{\odot}$
leads to a best fit to the data, which is shown in Figure~6.  This result
and values measured for the other stars are listed in Table~4.

     The upper panel of Figure~6 shows the same best fit as in the bottom
panel, but explicitly shows the contribution of the continuum to the total
flux of the model spectrum, and also shows the contributions of higher
orders 2--5 to the total model spectrum, illustrating that these higher
orders can contribute a significant amount of flux at certain wavelengths.
The best-fit model spectra for all of our stars are shown in Figure~3,
assuming our best estimates for Fe/H.  The emission measure distributions
shown in Figure~5 are normalized based on these absolute abundance
determinations.

     The EM(T) curves of the five K dwarfs are very similar, both in their
shape and in their magnitude, emphasizing once again the similarity of
these stars and their coronae.  All show a dramatic increase in emission
measure from $\log T=5.9$ to $\log T=6.1$.  The existence of this ``cliff''
relies entirely on measurements (or upper limits) of the Fe~IX $\lambda$171
line and a few Ne~VIII lines (see Table~3), which are the only lines
formed at temperatures low enough to constrain EM(T) below $\log T=6.0$.
All the EM(T) curves in Figure~5 also have peaks near $\log T=6.5-6.6$,
while at higher temperatures the emission measures decrease.  Based on
these results, we conclude that there is no evidence from these data that
the coronae of the five K dwarfs have any substantial differences in
temperature distribution, though small variations are present,
as suggested by the different O~VIII/O~VII ratios for 36~Oph~A and B,
for example.

     The only clear signature of coronal differences among these stars
is from the abundance measurements.  In Figure~7, the coronal abundances
listed in Table~4 are plotted relative to the stellar photospheric
abundances listed in Table~1.  Note that these photospheric abundances,
which are quoted relative to solar abundances, are measured by direct
line-by-line comparison of solar and stellar absorption lines,
meaning that the abundances are truly relative abundances that do
not assume any absolute abundance measurements for the Sun or the stars
\citep{cap04}.  We use the \citet{ng98}
solar photospheric abundances listed in Table~4 to convert the
relative abundances in Table~1 to absolute reference photospheric
abundances, for use in the creation of Figure~7.  There are no
photospheric abundances for 36~Oph~B or 70~Oph~B (see Table~1), so we
simply assume that these stars have abundances identical to their companion
stars, 36~Oph~A and 70~Oph~A, respectively.  Also, there are no stellar
measurements of N, Ne, or S abundances for any of the stars, so we simply
assume that the relative O abundances quoted in Table~1 apply to these
other 3 elements.

     The abundances in Figure~7 are plotted versus FIP, where
a dotted line is used to divide the elements into low-FIP
and high-FIP regimes.  A few of the stars seem to show some evidence for
a solar-like FIP effect (e.g., 70~Oph A, 36~Oph~AB), with the low-FIP
elements being higher in abundance relative to the high-FIP elements.
However, neon seems to be discrepant, with higher abundances
than the other high-FIP elements.  Another curious aspect of
these measurements is that even for those stars that show the solar-like
FIP bias (i.e., high low-FIP abundances), the measurements are
different from the standard solar-like FIP bias in that they seem to be
more consistent with the high-FIP elements being depleted in the corona
rather than the low-FIP elements being enhanced.

     The apparent lack of any evident low-FIP abundance enhancements in
the corona for any of our stars originates from our measurements of the
absolute Fe/H abundances using the line-to-continuum ratios (see
Fig.~6).  These Fe/H measurements range from
${\rm [Fe/H]}=0.68-1.57{\rm [Fe/H]_{*}}$, where ${\rm [Fe/H]_{*}}$ is the
reference photospheric abundance.  Only 70~Oph~A actually
has a ratio above the stellar photospheric value (see Fig.~7).  Similar
Fe/H values are measured by \citet{at05} from
XMM spectra of six solar-like G stars, and many of these stars
show evidence for high-FIP depletions in the corona similar to those
seen for 70~Oph~A and 36~Oph~AB in Figure~7.  Some analyses of solar
flare X-ray spectra have also found similar abundance patterns
\citep{af95}.  These results seem to suggest that
stellar fractionation processes are capable of both depleting high-FIP
elements and enhancing low-FIP elements, depending on the circumstances.
These issues will be discussed more in \S5.2.

     Figure~8 puts the abundances from Figure~7 into the same plot for
all five stars, normalized to the Fe abundance.  Dotted lines connect the
high-FIP elements in the figure, indicating the high-FIP abundance
relative to Fe.  The different levels of FIP bias noted on the basis of
the line flux ratios (see \S4.3) are here apparent from the actual
abundance measurements, with Figure~8 suggesting the following sequence
of increasing FIP effect:  70~Oph~B, $\epsilon$~Eri, 36~Oph~A, 36~Oph~B,
70~Oph~A.  As noted in \S4.3, it is remarkable that the two 70~Oph stars
are on opposite ends of this sequence, with 70~Oph~A having a much
stronger FIP effect than 70~Oph~B despite the similarities between these
stars in almost all other respects.  Although the overall level of
FIP bias is different for the five stars, Figure~8 suggests similar
{\em relative} abundances for the high-FIP elements for all five stars, with
O being low, Ne being high, and C and N being generally somewhere in
between.

     Confidence in a true commonality for this high-FIP abundance pattern
suffers from the size of the error bars in the abundance measurements,
particularly for C and N (see Fig.~7), but high Ne/O ratios have been
found to exist for many stellar coronae \citep{jjd05}.  We list
Ne/O ratios for our stars in Table~4.  These ratios are all about a
factor of 2 larger than the solar value of ${\rm Ne/O}=0.18$
\citep{ng98,jts05,pry05}.
The average and standard deviation of the K dwarf Ne/O ratios are
${\rm Ne/O}=0.37\pm 0.04$, which agrees well with the average value of
${\rm Ne/O}=0.41$ derived by \citet{jjd05} from a much larger
sample of stars.  The \citet{jjd05} sample consists almost
entirely of very active stars.  That our sample yields similar
results demonstrates that the high coronal Ne/O ratios extend to less
active stars, though it must be emphasized that our sample of K dwarfs
are still significantly more active than the comparatively inactive Sun.
Coronal abundance measurements for stars as inactive as the Sun are
difficult, since their faintness in X-rays makes it difficult to
obtain X-ray spectra with sufficient S/N to measure precise abundances.

\section{DISCUSSION}

\subsection{The 70 Oph Abundance Discrepancy}

     Perhaps the most interesting result of our analysis is the coronal
abundance difference between the two members of the 70~Oph binary, which
is unexpected since these two stars are so similar in all other respects.
Previous stellar observations have shown that the traditional
solar-like FIP effect is generally observed for stars of low to moderate
activity
\citep{jml96,jjd97,jml99}. 
In contrast, for very active stars an inverse FIP effect is generally
observed, where low-FIP elements have coronal abundances that are
{\em lower} relative to photospheric values than is the case for high-FIP
elements
\citep{acb01,jjd01,mg01,dph01,jsf03} 
Coronae of stars of intermediate activity are generally found to
have abundances with little or no dependence on FIP
\citep{jjd95,ma01,bb05}. 
Thus, the general picture is of the FIP effect being dependent on activity,
with the low-FIP element abundances decreasing in the corona with
increasing activity
\citep{ma03,at05}. 

     {\em However, if two stars as similar as 70~Oph~A and B can have
drastically different levels of FIP bias, this makes it very difficult
for any simple paradigm to explain why different stars have different
coronal abundance patterns.}  One potential way out of this dilemma is
to appeal to time dependent effects to explain the curious 70~Oph
results.  Time variations in FIP bias are observed in solar active
regions, which emerge with nearly photospheric abundances and then
acquire a FIP bias that increases on a timescale of days
\citep{kgw01}.  Perhaps at the
time of our {\em Chandra} observation of 70~Oph, the visible part of
70~Oph~B's corona was dominated by young active regions while the visible
part of 70~Oph~A's corona was dominated by old active regions, in which
case the observed difference in FIP behavior is temporary.  The only
way to test this hypothesis would be to reobserve 70~Oph, preferably
several times.  \citet{jsf03} looked for abundance
variations in multiple observations of the very active K dwarf AB~Dor,
with the observations encompassing a range of activity levels for the
star.  However, no abundance variability was found.

\subsection{Depletions of High-FIP Elements in Stellar Coronae}

     Another puzzling result to arise from our coronal abundance
analysis is the apparent depletion of high-FIP elements in the
corona with respect to the photosphere, at least for a few of our
stars.  Although the origin of the FIP effect and inverse FIP effect is
not well understood, the cause is thought to lie in the chromosphere,
where the low-FIP elements are ionized and therefore subject to forces
that the neutral high-FIP elements are not subject to.  The general
assumption is that these forces, whatever their nature, preferentially
act on low-FIP elements, which leads to these elements having
nonphotospheric abundances in the corona.  For the Sun and relatively
inactive stars, the low-FIP coronal abundance trend is towards high
abundances, while for very active stars the trend is towards low
abundances.

     In this simple scenario, the elements with the highest FIP should
always have coronal abundances closest to the photospheric values since
they are less subject to whatever forces are causing the FIP effect.
This, however, is not consistent with what we see in our data.  For
example, for every single one of our stars the coronal O abundance is
farther from the photospheric abundance than the Fe and Si abundances,
with the data suggesting a depletion of O in the corona in every
case.   Note that measured photospheric O, Si, and Fe abundances exist
for our stars and are taken into account in the analysis (see Table~1),
so a lack of photospheric reference abundances is not an issue here,
though it is often a problem in interpreting the coronal abundance
measurements for very active stars \citep[see, e.g.,][]{jsf04}.
\citet{at05} find similar evidence for high-FIP
element depletions for many of the active solar-like stars in their
sample.  Other examples of coronal abundance patterns that are also
difficult for any simple FIP-based fractionation model to explain are
reported by \citet{acb01}, \citet{jjd01}, \citet{dph03}, and \citet{jsf04}.

     Another observational result that the simple FIP-dependent scenarios
cannot easily explain is that the Ne/O ratios of many different stars are
the same, regardless of whether the stars generally show a solar-like FIP
effect, no FIP effect, or an inverse FIP effect.  All of our stars and all
of those in the larger sample of \citet{jjd05} suggest
${\rm Ne/O}\approx 0.4$.  Either the processes that cause the
observed coronal abundance anomales are not dependent on FIP in any
simple, monotonic fashion, or there are other fractionation forces at
work that have nothing to do with FIP whatsoever.

     Some solar flare observations have also suggested abundance patterns
similar to that seen for 70~Oph~A in Figure~8, with high-FIP elements
being depleted in the corona \citep{njv81,af95}.
The Ne/O ratios observed from solar active regions
also show substantial variation about the mean solar value of 0.18,
demonstrating a deviation from the average FIP effect even in relatively
quiescent plasma \citep{jts96}.  It has been proposed that
photoionization by coronal X-rays can lead to nonequilibrium ionization
conditions in the atmospheric regions where fractionation is taking place
\citep{as91}.  Since Ne, for example, has a higher photoionization cross
section than O, Ne could be photionized while O remains neutral despite
Ne having a higher FIP.  Analogous effects have been proposed to explain
a factor of 2 depletion of He in the solar wind \citep{rvs89,rvs00}.

     For active stars, the irradiation of stellar
atmospheres by coronal X-rays would be more intense, potentially making
such photoionization effects more important.  This could perhaps explain
the high Ne/O ratios seen for active stars, but it does not obviously
explain other aspects of the observed abundance patterns, such as why
high-FIP elements can be depleted in stellar coronae.  Clearly, we are
still far from understanding the fractionation forces operating in
stellar atmospheres, and far from understanding why they appear to
operate so differently on different stars.

\subsection{The Ne/O Controversy}

     The whole issue of Ne/O ratios deserves additional comment
since solar/stellar Ne and O abundances have recently become
controversial topics.  The discord began when complex 3D hydrodynamic
models of the solar atmosphere were used to reassess solar abundances,
and it was then claimed that the abundances of many light elements such
as C, N, and O needed to be revised downward by $25-35$\%
\citep{cap01,cap02,ma05}.  Unfortunately, these suggested
changes seriously damage the remarkably good agreement
that exists between models of the solar interior and various
helioseismological measurements, when older abundances are used
\citep{sb04,jnb05a}.

     It has been noted that these problems can be mitigated by
increasing the assumed solar Ne abundance by a factor of 2--3
\citep{hma05,jnb05b}.  The reason why
such an increase can be reasonably proposed is that the solar Ne abundance
is poorly known.  There are no spectroscopic diagnostics of Ne in the
photosphere, which means that the solar Ne abundance must be inferred
from coronal or transition region measurements.  But this
is dangerous given the existence of fractionation processes that
result in differences between coronal and photospheric abundances
(see \S5.2).

     When \citet{jjd05} demonstrated that stellar Ne/O ratios
have values consistent with ${\rm Ne/O}=0.41$, which is
significantly higher than the ratio assumed for the Sun
(${\rm Ne/O}=0.18$), they
proposed that the stellar measurements represent a better estimate
of the solar/stellar photospheric Ne/O ratios than the lower solar
ratio.  Not only would this explain the remarkably consistent stellar
Ne/O ratios, but it would also solve the discrepancies between solar
interior models and helioseismology induced by the revised C, N, and O
abundances.  Unfortunately, very recent reassessments of the solar
coronal Ne/O ratio based on both full-disk and spatially resolved
data seem to confirm that the solar Ne/O ratio is truly a factor
of 2 lower than those observed from the stars in the \citet{jjd05}
sample, which are more active than the Sun \citep{pry05,jts05}.
Thus, the evidence at present seems to suggest
that the Sun really does have a different coronal Ne/O ratio than
active stars.

     Furthermore, there are measurements of Ne/O for stars
as inactive as the Sun that suggest low, solar-like ratios.  These
measurements generally carry large uncertainties because the
Ne~IX and Ne~X lines used to provide Ne abundances are generally
weak for the cooler coronae of inactive stars.  Nevertheless,
\citet{ar02} quote a value of ${\rm Ne/O}=0.21$ for Procyon, though
\citet{jsf04} find a higher value of ${\rm Ne/O}=0.40$.  \citet{ar03}
report ${\rm Ne/O}=0.18$ and ${\rm Ne/O}=0.24$ for
$\alpha$~Cen~A and B, respectively, and
\citet{at05} find ${\rm Ne/O}\approx 0.14$ for $\beta$~Com, though
with large uncertainties.  A compilation of measurements in
\citet{mg04}, including those mentioned above, provides some
evidence for a weak correlation of Ne/O with activity.

     As mentioned in \S4.4, the high Ne/O ratios seen for our K dwarfs
are significant because they demonstrate that the high Ne/O ratio found
by \citet{jjd05} applies to stars less active than most of those
in their sample.  But our K dwarfs are still significantly more active
than the Sun.  There are two possible explanations for the different
Ne/O ratios that exist for the Sun and more active stars:  (a) The
stellar Ne/O measurements are representative of the true cosmic abundance
and the low solar Ne/O measurements are a product of some fractionation
mechanism operating in the solar atmosphere but not the active stars, or
(b) The solar Ne/O measurement is characteristic of the true cosmic
abundance and the high stellar ratios are due to a fractionation mechanism
that is not present for the Sun.  At first glance, the latter
interpretation seems more likely, partly because the low solar Ne/O
abundances are known to exist down to at least transition region
temperatures \citep{pry05}, and partly because it is easier to imagine
active stars having strong fractionation effects that result in
substantial differences between coronal and photospheric Ne/O ratios.
However, this would leave the solar interior problems unresolved.

     Without direct photospheric measurements of Ne, it is not possible
to say much about solar/stellar Ne abundances with any great amount
of confidence.  Perhaps for Ne it is best to measure the cosmic abundance
from a very different astronomical source.  We note that \citet{gg04}
have estimated O and Ne abundances in the local ISM using
{\em Ulysses} and {\em ACE} measurements of pickup ions within our
solar system, and they find ${\rm Ne/O}\approx 0.41$.  This is in
better agreement with the active star ratio than the solar ratio,
though it must be noted that the ISM measurements come with their own
set of systematic uncertainties and assumptions regarding dust
depletion, deflection at the heliopause, and ISM ionization fraction.

\subsection{Do Winds Correlate with Coronal FIP Bias?}

     A major motivation behind our {\em Chandra} observations
was to see whether coronal differences in our K dwarf sample could be
correlated with observed differences in the strengths of their winds.
However, the only coronal properties that clearly vary in this sample
of stars are the abundances.  The overall coronal activity levels
of these stars are about the same, and the {\em Chandra} data cannot
distinguish any significant density or temperature differences in the
stellar coronae.

     Can the observed abundance variations be correlated with wind
strength?  The mass loss rates per unit surface area of 36~Oph,
$\epsilon$~Eri, and 70~Oph (in solar units) are 18.1, 49.3, and
85.7, respectively (see Table~1).  (Unfortunately, we only know the
combined mass loss rates for the 36~Oph and 70~Oph binaries and not the
contributions of the individual stars.)  The 36~Oph binary has the
weakest wind, and both stars have coronae that show a modest FIP effect
(see Fig.~8).  This suggests a potential connection between weak winds
and strong FIP bias.  The stronger wind and weak-or-no FIP effect of
$\epsilon$~Eri would support such a connection, but the fact that
70~Oph consists of both a high FIP-bias star (70~Oph~A) and a
no-or-inverse FIP-bias star (70~Oph~B) complicates the interpretation
of its high mass loss measurement.
If the FIP-bias/wind anticorrelation suggested by 36~Oph and
$\epsilon$~Eri is true, that would imply that 70~Oph~B must be
responsible for most of 70~Oph's strong wind.  Small number statistics
preclude any definitive conclusions on the nature of any FIP/wind
connection, but we note that for the Sun different levels of FIP bias
are seen for low and high speed streams \citep{rvs00}.
This suggests that a FIP/wind connection of some sort is not implausible,
but it must be noted that our X-ray spectra will be characteristic of
brighter, high density plasma still confined in coronal loops rather
than outflowing wind material, so any FIP/wind connection would have
to be indirect.

\section{SUMMARY}

     We have analyzed {\em Chandra} LETG/HRC-S spectra of five very
similar moderately active K dwarf stars (36~Oph~AB, 70~Oph~AB, and
$\epsilon$~Eri), measuring emission measure distributions and coronal
abundances from the spectra for all five stars.  Our results are
summarized as follows:
\begin{description}
\item[1.] Our observations of 36~Oph and 70~Oph have resolved these
  binaries in X-rays for the first time, showing that 36~Oph~A and
  70~Oph~A account for 58.2\% and 60.4\% of the binaries' X-ray
  emission, respectively.
\item[2.] We estimate coronal densities from the flux ratios of the
  O~VII $\lambda$21.8 triplet lines.  All five stars have
  densities consistent with $\log n_{e}\approx 10.0$.
\item[3.] The emission measure distributions of our five K dwarfs are
  very similar, so there is no evidence for significant coronal
  temperature variations among these stars.
\item[4.] Elemental abundances are the only coronal properties that
  clearly are different within our sample of moderately active K
  dwarfs.  The two extremes are 70~Oph~A, which shows a prominent
  solar-like FIP effect, and 70~Oph~B, which has no FIP bias
  at all or possibly even a weak inverse FIP effect.  The strong
  abundance difference between 70~Oph~A and 70~Oph~B is very surprising,
  considering how similar these two stars are in almost every other
  respect.  This will be very difficult for any simple model of the
  FIP effect to explain, though we cannot rule out the possibility
  that the apparent 70~Oph~AB discrepancy could be a temporary product
  of coronal abundance variability, such as that seen during the
  evolution of solar active regions.
\item[5.] For our stars, high-FIP elements generally seem to be depleted
  in the corona relative to the photosphere, which is very different
  from the standard picture of the solar FIP effect, with low-FIP elements
  enhanced in the corona and high-FIP elements unchanged from
  photospheric abundances.  It is very unlikely that any simple
  fractionation model based on FIP as the sole discriminant is
  likely to explain the variety of abundance patterns that have
  been observed in stellar coronae.
\item[6.] The Ne/O ratio is essentially the same for our five stars,
  ${\rm Ne/O}=0.37\pm 0.04$.  This value is a factor of 2 higher
  than the solar value but it agrees very well with the measurement
  of ${\rm Ne/O}=0.41$ from \citet{jjd05} based on a
  large selection of stars.  Our stellar sample is generally less
  active than the \citet{jjd05} sample, so our measurements
  demonstrate that high stellar Ne/O coronal abundance ratios extend to
  lower activity levels.
\item[7.] A small sample size precludes any definitive conclusions
  about whether the coronal abundance variations we see in our data
  are connected in any way with variations that also exist in
  measured mass loss rates for these stars.  The 36~Oph~AB and
  $\epsilon$~Eri measurements could suggest an anticorrelation between
  coronal FIP bias and wind strength, which would mean that 70~Oph~B
  would have to be responsible for the particularly strong wind
  observed from the 70~Oph binary.  Unfortunately, this prediction
  is untestable at the present time.
\end{description}

\acknowledgments

     We would like to thank the referee, Manuel G\"{u}del, for
helpful comments.
Support for this work was provided by the National Aeronautics and Space
Administration through Chandra Award Number GO4-5170X issued by the Chandra
X-ray Observatory Center, which is operated by the Smithsonian
Astrophysical Observatory for and on behalf of the National Aeronautics
and Space Administration under contract NAS8-03060.

\clearpage

\begin{deluxetable}{ccccccc}
\tabletypesize{\scriptsize}
\tablecaption{Stellar Information}
\tablecolumns{7}
\tablewidth{0pt}
\tablehead{
  \colhead{Property\tablenotemark{a}} & \multicolumn{2}{c}{36 Ophiuchi} &
    \multicolumn{2}{c}{70 Ophiuchi} & \colhead{$\epsilon$ Eridani} &
    \colhead{Refs.} \\
  \colhead{} & \colhead{A} & \colhead{B} & \colhead{A} & \colhead{B} &
    \colhead{} & \colhead{}}
\startdata
Other Name   & HD 155886 & HD 155885 & HD 165341 & HD 165341B& HD 22049 & \\
Spect.\ Type &  K1 V     &   K1 V    &   K0 V    &   K5 V    &   K1 V   & \\
Dist.\ (pc)  &  5.99     &   5.99    &   5.09    &   5.09    &   3.22   & 1\\
Radius (R$_{\odot}$)&0.69&   0.59    &   0.85    &   0.66    &   0.78   & 2\\
P$_{rot}$ (days)& 20.3   &   22.9    &   19.7    &   22.9    &   11.7  &3,4\\
Log L$_{x}$\tablenotemark{b} & 28.10 & 27.96 & 28.27 & 28.09 & 28.32    & 5 \\
$\dot{M}$ ($\dot{M}_{\odot}$)\tablenotemark{c} & \multicolumn{2}{c}{15} &
  \multicolumn{2}{c}{100} & 30 & 6 \\
$\log N_{H}$\tablenotemark{d} & \multicolumn{2}{c}{17.85} &
  \multicolumn{2}{c}{18.06} & 17.88 & 7 \\
$\log {\rm [C/C_{\odot}]}$  & $-0.40$ &   ...  & $-0.10$ &
   ...  & $-0.24$ & 8 \\
$\log {\rm [O/O_{\odot}]}$  & $-0.14$ &   ...  & $0.03$  &
   ...  & $-0.04$ & 8 \\
$\log {\rm [Mg/Mg_{\odot}]}$ & $-0.28$ &   ...  & $0.09$  &
   ...  & $-0.03$ & 8 \\
$\log {\rm [Si/Si_{\odot}]}$ & $-0.07$ &   ...  & $0.18$  &
   ...  & $-0.01$ & 8 \\
$\log {\rm [Ca/Ca_{\odot}]}$ & $-0.15$ &   ...  & $0.09$  &
   ...  & $-0.01$ & 8 \\
$\log {\rm [Fe/Fe_{\odot}]}$ & $-0.30$ &   ...  & $-0.05$ &
   ...  & $-0.06$ & 8 \\
$\log {\rm [Ni/Ni_{\odot}]}$ & $-0.20$ &   ...  & $0.06$  &
   ...  & $-0.06$ & 8 \\
\enddata
\tablenotetext{a}{The quantities in square brackets are stellar
  photospheric abundances relative to solar values.}
\tablenotetext{b}{X-ray luminosities (ergs~s$^{-1}$) from ROSAT all-sky
  survey data, where the results of this paper are used to establish the
  contributions of the individual stars for the unresolved 36~Oph and
  70~Oph binaries.}
\tablenotetext{c}{Mass loss rate measurements from astrospheric absorption
  detections, where for 36~Oph and 70~Oph the measurement is for the
  combined mass loss from both stars of the binary.}
\tablenotetext{d}{Interstellar H~I column density.}
\tablerefs{(1) Perryman et al.\ 1997. (2) Barnes et al.\ 1978. (3) Donahue
  et al.\ 1996. (4) Noyes et al.\ 1984. (5) Schmitt \& Liefke 2004.
  (6) Wood et al.\ 2005a. (7) Wood et al.\ 2005b. (8) Allende Prieto
  et al.\ 2004.}
\end{deluxetable}

\begin{deluxetable}{ccccc}
\tabletypesize{\scriptsize}
\tablecaption{{\em Chandra} LETG/HRC-S Observations}
\tablecolumns{5}
\tablewidth{0pt}
\tablehead{
  \colhead{Star} & \colhead{Obs.\ ID} & \colhead{Date} &
    \colhead{Start Time} & \colhead{Exp.\ Time} \\
  \colhead{} & \colhead{} & \colhead{} & \colhead{(UT)} & \colhead{(ksec)}}
\startdata
$\epsilon$~Eri & 1869 & 2001 Mar.\ 21&  7:17:12 &105.91 \\
36~Oph         & 4483 & 2004 June 1  & 10:22:33 & 77.90 \\
70~Oph         & 4482 & 2004 July 19 & 22:59:01 & 78.94 \\
\enddata
\end{deluxetable}

\begin{deluxetable}{lccccccc}
\tabletypesize{\scriptsize}
\tablecaption{{\em Chandra} Line Measurements}
\tablecolumns{8}
\tablewidth{0pt}
\tablehead{
  \colhead{Ion} & \colhead{$\lambda_{rest}$} & \colhead{$\log T$} &
    \multicolumn{5}{c}{Counts} \\
  \colhead{} & \colhead{(\AA)} & \colhead{} &
    \colhead{$\epsilon$ Eri} & \colhead{70 Oph A} &
    \colhead{70 Oph B} & \colhead{36 Oph A} & \colhead{36 Oph B}}
\startdata
Si XIII  &   6.648 &6.99 &$ 127.1\pm 22.4$ &$  50.3\pm 19.8$ &
  $  22.5\pm 18.8$ &         $<32.4$ &         $<32.0$ \\
Si XIII  &   6.688 &6.99 &                 &                 &
                   &                 &                 \\
Si XIII  &   6.740 &6.99 &                 &                 &
                   &                 &                 \\
Mg XII   &   8.419 &7.11 &$  51.9\pm 15.7$ &         $<32.0$ &
           $<32.8$ &         $<32.4$ &         $<30.8$ \\
Mg XII   &   8.425 &7.11 &                 &                 &
                   &                 &                 \\
Mg XI    &   9.169 &6.80 &$ 185.4\pm 29.6$ &         $<38.8$ &
           $<40.0$ &$  38.9\pm 21.1$ &$  25.7\pm 20.4$ \\
Mg XI    &   9.231 &6.80 &                 &                 &
                   &                 &                 \\
Mg XI    &   9.314 &6.79 &                 &                 &
                   &                 &                 \\
Ne IX    &  11.547 &6.61 &$  61.8\pm 17.7$ &         $<34.0$ &
           $<33.0$ &         $<31.6$ &         $<32.4$ \\
Ne X     &  12.132 &6.87 &$ 481.8\pm 32.5$ &$  58.3\pm 17.4$ &
  $  73.7\pm 16.3$ &$  23.8\pm 13.9$ &         $<32.0$ \\
Ne X     &  12.138 &6.87 &                 &                 &
                   &                 &                 \\
Ne IX    &  13.447 &6.58 &$ 818.1\pm 39.8$ &$  85.6\pm 21.5$ &
  $  92.4\pm 18.8$ &$  94.0\pm 21.9$ &$  40.7\pm 16.0$ \\
Ne IX    &  13.553 &6.58 &                 &                 &
                   &                 &                 \\
Ne IX    &  13.699 &6.58 &$ 367.8\pm 30.4$ &$  30.2\pm 16.5$ &
  $  69.4\pm 17.9$ &$  38.1\pm 15.9$ &$  29.2\pm 15.5$ \\
Fe XVIII &  14.203 &6.74 &$ 205.1\pm 31.4$ &$  39.4\pm 15.6$ &
           $<33.8$ &         $<31.6$ &         $<32.0$ \\
Fe XVIII &  14.208 &6.74 &                 &                 &
                   &                 &                 \\
Fe XVII  &  15.015 &6.59 &$1207.1\pm 46.8$ &$ 231.7\pm 24.6$ &
  $  83.7\pm 18.3$ &$ 103.2\pm 18.3$ &$  67.7\pm 18.5$ \\
Fe XVII&  15.262 &6.59 &$ 709.1\pm 42.3$ &$ 155.9\pm 22.8$ &
  $  69.8\pm 19.5$ &$  61.8\pm 17.0$ &$  40.9\pm 17.5$ \\
O VIII &  15.176 &6.65 &                 &                 &
                   &                 &                 \\
Fe XIX &  15.198 &6.83 &                 &                 &
                   &                 &                 \\
O VIII &  16.006 &6.63 &$ 515.5\pm 35.0$ &$  86.7\pm 20.3$ &
  $  64.8\pm 18.9$ &$  59.6\pm 18.2$ &$  50.5\pm 17.6$ \\
Fe XVIII& 16.005 &6.73 &                 &                 &
                   &                 &                 \\
Fe XVIII& 16.072 &6.73 &                 &                 &
                   &                 &                 \\
Fe XVII  &  16.778 &6.58 &$ 721.7\pm 35.7$ &$ 165.5\pm 22.4$ &
  $  62.9\pm 18.9$ &$  46.1\pm 14.8$ &$  52.0\pm 17.9$ \\
Fe XVII  &  17.053 &6.58 &$1660.9\pm 50.6$ &$ 325.2\pm 27.5$ &
  $ 128.4\pm 20.9$ &$ 138.4\pm 22.1$ &$ 100.2\pm 19.5$ \\
Fe XVII  &  17.098 &6.58 &                 &                 &
                   &                 &                 \\
O VII    &  18.627 &6.34 &$ 152.9\pm 21.8$ &$  42.5\pm 17.9$ &
           $<33.6$ &$  18.1\pm 12.9$ &$  19.9\pm 15.1$ \\
O VIII   &  18.967 &6.59 &$2272.3\pm 55.7$ &$ 283.7\pm 25.2$ &
  $ 320.0\pm 27.1$ &$ 241.3\pm 23.6$ &$ 154.2\pm 21.7$ \\
O VIII   &  18.973 &6.59 &                 &                 &
                   &                 &                 \\
O VII    &  21.602 &6.32 &$ 773.2\pm 35.7$ &$ 128.6\pm 19.2$ &
  $ 119.4\pm 19.2$ &$  82.4\pm 17.1$ &$  96.9\pm 20.1$ \\
O VII    &  21.807 &6.32 &$ 152.8\pm 21.6$ &$  25.5\pm 13.7$ &
  $  15.0\pm 13.5$ &$  29.2\pm 13.2$ &$  25.6\pm 13.2$ \\
O VII    &  22.101 &6.31 &$ 504.6\pm 29.9$ &$  74.7\pm 17.5$ &
  $  44.0\pm 14.9$ &$  56.7\pm 16.3$ &$  74.5\pm 17.2$ \\
N VII    &  24.779 &6.43 &$ 211.7\pm 26.5$ &$  34.8\pm 14.3$ &
  $  49.8\pm 18.1$ &$  22.7\pm 13.3$ &$  20.9\pm 14.5$ \\
N VII    &  24.785 &6.43 &                 &                 &
                   &                 &                 \\
C VI     &  33.734 &6.24 &$ 340.5\pm 25.4$ &$  62.1\pm 17.3$ &
  $  71.1\pm 19.1$ &$  27.7\pm 13.6$ &$  18.1\pm 14.7$ \\
C VI     &  33.740 &6.24 &                 &                 &
                   &                 &                 \\
S XIII   &  35.667 &6.43 &$  94.1\pm 21.5$ &         $<33.2$ &
           $<33.2$ &         $<30.8$ &         $<32.0$ \\
Si XI    &  43.763 &6.25 &$ 133.2\pm 22.8$ &$  41.0\pm 15.1$ &
           $<33.0$ &$  19.6\pm 14.6$ &         $<30.4$ \\
Si XII   &  44.019 &6.44 &$ 185.2\pm 24.0$ &$  52.4\pm 17.2$ &
  $  34.4\pm 14.3$ &$  18.8\pm 12.9$ &         $<31.4$ \\
Si XII   &  44.165 &6.44 &$ 288.0\pm 26.7$ &$  74.9\pm 17.8$ &
  $  37.0\pm 17.3$ &$  28.9\pm 15.1$ &$  27.0\pm 14.9$ \\
Si XI    &  49.222 &6.24 &$ 195.8\pm 26.1$ &$  50.7\pm 16.7$ &
           $<33.2$ &$  35.2\pm 15.4$ &$  42.0\pm 17.0$ \\
Fe XVI   &  50.361 &6.43 &$ 246.2\pm 28.0$ &$  40.0\pm 16.1$ &
  $  28.1\pm 16.9$ &         $<33.8$ &         $<32.8$ \\
Fe XVI &  50.565 &6.43 &$ 147.3\pm 23.6$ &$  53.1\pm 16.8$ &
           $<31.6$ &         $<31.8$ &         $<30.8$ \\
Si X   &  50.524 &6.15 &                 &                 &
                   &                 &                 \\
Si X     &  50.691 &6.15 &$  71.7\pm 21.2$ &$  40.0\pm 15.9$ &
           $<30.0$ &         $<31.0$ &         $<31.6$ \\
Fe XV    &  52.911 &6.32 &$  50.1\pm 16.0$ &         $<27.2$ &
           $<28.6$ &         $<26.4$ &         $<26.0$ \\
Fe XVI   &  54.127 &6.43 &$  96.7\pm 17.3$ &         $<28.2$ &
           $<27.8$ &         $<26.8$ &         $<27.6$ \\
Fe XVI   &  54.710 &6.43 &$ 121.0\pm 19.4$ &$  32.8\pm 15.6$ &
           $<32.2$ &         $<27.8$ &         $<27.8$ \\
Mg X     &  57.876 &6.22 &$ 144.9\pm 23.2$ &$  53.2\pm 15.0$ &
           $<31.6$ &         $<31.2$ &$  19.7\pm 13.0$ \\
Mg X     &  57.920 &6.22 &                 &                 &
                   &                 &                 \\
Fe XV    &  59.405 &6.32 &$ 107.8\pm 20.8$ &$  30.6\pm 14.5$ &
           $<32.8$ &         $<31.0$ &         $<30.2$ \\
Fe XVI   &  63.711 &6.42 &$ 140.6\pm 18.8$ &$  37.8\pm 12.5$ &
  $  19.3\pm 11.5$ &$  28.7\pm 13.1$ &$  20.7\pm 12.7$ \\
Fe XVI   &  66.357 &6.42 &$ 363.2\pm 29.1$ &$  88.7\pm 20.3$ &
  $  42.8\pm 17.6$ &$  38.8\pm 17.5$ &$  40.9\pm 17.6$ \\
Fe XVI   &  66.249 &6.42 &                 &                 &
                   &                 &                 \\
Fe XV    &  69.682 &6.32 &$ 285.2\pm 25.8$ &$  76.5\pm 17.4$ &
           $<31.4$ &$  22.3\pm 12.9$ &$  27.0\pm 13.2$ \\
Fe XV    &  73.472 &6.32 &$ 118.9\pm 20.1$ &         $<33.6$ &
           $<32.4$ &         $<29.4$ &         $<31.4$ \\
Ne VIII  &  88.082 &5.96 &$ 176.0\pm 21.3$ &$  21.2\pm 14.9$ &
  $  67.2\pm 19.2$ &$  24.5\pm 14.4$ &         $<28.0$ \\
Ne VIII  &  88.120 &5.96 &                 &                 &
                   &                 &                 \\
Fe XVIII &  93.923 &6.68 &$ 275.7\pm 33.0$ &$  62.3\pm 20.9$ &
  $  51.5\pm 20.5$ &         $<42.0$ &         $<42.0$ \\
Ne VIII  &  98.260 &5.94 &$ 153.7\pm 27.3$ &         $<43.4$ &
           $<42.2$ &         $<42.0$ &         $<41.8$ \\
Fe XVIII & 103.937 &6.68 &$ 108.1\pm 28.9$ &         $<42.8$ &
           $<42.6$ &         $<41.6$ &         $<39.6$ \\
Fe XIX   & 108.355 &6.79 &$ 112.0\pm 30.1$ &$  46.9\pm 22.6$ &
  $  33.1\pm 19.8$ &         $<41.6$ &         $<42.6$ \\
Fe XX    & 121.845 &6.88 &$  55.9\pm 26.7$ &         $<40.4$ &
           $<41.0$ &         $<41.4$ &         $<39.0$ \\
Ca XII   & 141.038 &6.25 &$  53.7\pm 24.2$ &         $<41.0$ &
           $<42.0$ &         $<41.0$ &         $<41.4$ \\
Ni XII   & 152.154 &6.23 &$  58.5\pm 24.3$ &         $<43.4$ &
           $<43.0$ &         $<42.4$ &         $<40.4$ \\
Fe IX    & 171.073 &5.95 &$ 114.2\pm 21.3$ &$  45.6\pm 18.2$ &
  $  25.2\pm 17.3$ &         $<35.4$ &$  18.2\pm 16.3$ \\
\enddata
\end{deluxetable}

\begin{deluxetable}{ccccccc}
\tabletypesize{\scriptsize}
\tablecaption{Densities and Abundances}
\tablecolumns{7}
\tablewidth{0pt}
\tablehead{
  \colhead{} & \colhead{$\epsilon$ Eri} & \colhead{70 Oph A} &
    \colhead{70 Oph B} & \colhead{36 Oph A} & \colhead{36 Oph B} &
    \colhead{Sun\tablenotemark{a}}}
\startdata
$\log n_{e}$    & $9.82^{+0.33}_{-0.72}$ & $<10.89$ & $<12.06$ &
  $10.55\pm 0.51$ & $<10.88$ & 9.4 \\
${\rm Fe/Fe_{\odot}}$ & 0.7 & 1.4 & 0.6 & 0.5 & 0.5 & 1 \\
$\log {\rm [Fe/H]}$  & $-4.65$ & $-4.35$ & $-4.72$&$-4.80$&$-4.80$&$-4.50$\\
$\log {\rm [C/Fe]}$  & $0.80_{-0.06}^{+0.04}$ & $0.48_{-0.15}^{+0.16}$ &
  $1.02_{-0.13}^{+0.39}$ & $0.53_{-0.35}^{+0.28}$ & $0.44_{-0.34}^{+0.44}$ &
  1.02 \\
$\log {\rm [N/Fe]}$  & $0.36_{-0.09}^{+0.04}$ & $0.14_{-0.36}^{+0.16}$ &
  $0.71_{-0.29}^{+0.25}$ & $0.33_{-0.56}^{+0.26}$ & $0.21_{-0.55}^{+0.38}$ &
  0.42 \\
$\log {\rm [O/Fe]}$  & $1.11_{-0.03}^{+0.02}$ & $0.82_{-0.05}^{+0.10}$ &
  $1.33_{-0.07}^{+0.16}$ & $1.08_{-0.10}^{+0.16}$ & $1.09_{-0.16}^{+0.15}$ &
  1.33 \\
$\log {\rm [Ne/Fe]}$ & $0.67_{-0.02}^{+0.04}$ & $0.38_{-0.12}^{+0.14}$ &
  $0.94_{-0.11}^{+0.12}$ & $0.69_{-0.24}^{+0.14}$ & $0.58_{-0.64}^{+0.21}$ &
  0.58 \\
$\log {\rm [Mg/Fe]}$ & $0.11_{-0.07}^{+0.07}$ & $0.16_{-0.17}^{+0.15}$ &
  ...                    & $0.47_{-1.03}^{+0.10}$ & $0.15_{-0.69}^{+0.40}$ &
  0.08 \\
$\log {\rm [Si/Fe]}$ &$-0.02_{-0.06}^{+0.06}$ & $0.01_{-0.18}^{+0.09}$ &
  $0.19_{-0.48}^{+0.24}$ & $0.11_{-0.38}^{+0.13}$ &$-0.07_{-0.56}^{+0.36}$ &
  0.05 \\
$\log {\rm [S/Fe]}$  &$-0.20_{-0.26}^{+0.14}$ & ...                    &
  ...                    & ...                    & ...                    &
  $-0.17$\\
$\log {\rm [Ca/Fe]}$ &$-0.65_{-1.01}^{+0.25}$ & ...                    &
  ...                    & ...                    & ...                    &
  $-1.14$\\
$\log {\rm [Ni/Fe]}$ &$-1.37_{-0.62}^{+0.19}$ & ...                    &
  ...                    & ...                    & ...                    &
  $-1.25$\\
${\rm Ne/O}$       &  0.36                  &  0.36                  &
  0.40                   &  0.41                  &  0.31                  &
  0.18   \\
\enddata
\tablenotetext{a}{The coronal density quoted for the Sun is from
  McKenzie (1987), while the abundances listed are photospheric abundances
  from Grevesse \& Sauval (1998).}
\end{deluxetable}


\begin{deluxetable}{lccccc}
\tabletypesize{\scriptsize}
\tablecaption{{\em HST} Line Measurements}
\tablecolumns{6}
\tablewidth{0pt}
\tablehead{
  \colhead{Ion} & \colhead{$\lambda_{rest}$} & \colhead{$\log T$} &
    \multicolumn{3}{c}{Flux ($10^{-15}$ ergs cm$^{-2}$ s$^{-1}$)} \\
  \colhead{} & \colhead{(\AA)} & \colhead{} &
    \colhead{$\epsilon$ Eri} & \colhead{70 Oph A} &
    \colhead{36 Oph A}}
\startdata
C III    &1174.933 &4.80 &$  41.0\pm 6.0$ &$  22.4\pm 1.7$ &$   4.2\pm 0.6$\\
C III    &1175.263 &4.80 &$  41.0\pm 6.0$ &$  15.9\pm 1.7$ &$   3.6\pm 0.7$\\
C IIIbl  &1175.69 &4.80 &$ 149.0\pm 6.0$ &$  70.5\pm 3.6$ &$  22.3\pm 1.0$\\
C III    &1175.987 &4.80 &$  41.0\pm 6.0$ &$  21.0\pm 2.1$ &$   3.6\pm 0.8$\\
C III    &1176.370 &4.80 &$  46.0\pm 6.0$ &$  24.6\pm 1.9$ &$   4.9\pm 0.9$\\
Si II    &1190.416 &4.37 &$  10.0\pm 3.0$ &$   5.8\pm 0.7$ &$   2.1\pm 0.4$\\
Si II    &1193.290 &4.37 &$   9.0\pm 4.0$ &$   4.2\pm 0.6$ &$   2.0\pm 0.4$\\
Si II    &1194.500 &4.37 &$  18.0\pm 3.0$ &$  10.6\pm 0.9$ &$   2.8\pm 0.3$\\
Si II    &1197.394 &4.37 &$  10.0\pm 4.0$ &$   3.6\pm 0.9$ &$   1.2\pm 0.3$\\
Si III   &1206.500 &4.60 &$ 360.0\pm 10.0$ &$ 257.9\pm 3.9$ &$ 62.8\pm 1.3$\\
H I      &1215.671 &4.20 &$ 48800\pm 9760$ &$21400\pm 4280$&$14200\pm 2840$\\
O V      &1218.344 &5.33 &$  68.0\pm 4.0$ &$  28.0\pm 2.0$ &$  13.8\pm 0.9$\\
N V      &1238.821 &5.25 &$  94.0\pm 3.0$ &$  45.1\pm 1.2$ &$  10.7\pm 0.5$\\
Fe XII   &1242.01 &6.15 &$  12.0\pm 0.8$ &$   6.7\pm 0.5$ &$   2.1\pm 0.2$\\
N V      &1242.804 &5.25 &$  46.0\pm 2.0$ &$  22.7\pm 0.7$ &$   5.3\pm 0.4$\\
S II     &1253.811 &4.43 &$   6.6\pm 0.5$ &$   3.0\pm 0.3$ &$   1.3\pm 0.2$\\
S II     &1259.519 &4.43 &$   9.9\pm 0.5$ &$   4.0\pm 0.4$ &$   1.2\pm 0.2$\\
Si II    &1260.422 &4.35 &$  19.1\pm 3.0$ &$  10.0\pm 0.5$ &$   3.7\pm 0.3$\\
Si II    &1264.738 &4.35 &$  50.5\pm 5.0$ &$  35.2\pm 0.8$ &$   9.7\pm 0.4$\\
Si II    &1265.002 &4.35 &$  17.2\pm 3.0$ &$   9.3\pm 0.6$ &$   3.1\pm 0.3$\\
C I      &1288.422 &4.25 &$   5.7\pm 0.4$ &$   2.8\pm 0.2$ &$   0.8\pm 0.1$\\
Si III   &1294.545 &4.59 &$   5.2\pm 0.5$ &$   3.4\pm 0.3$ &$   0.8\pm 0.2$\\
Si III   &1296.726 &4.59 &$   4.1\pm 0.5$ &$   2.4\pm 0.2$ &$   0.8\pm 0.2$\\
Si IIIbl &1298.93 &4.59 &$  16.7\pm 0.5$ &$  10.3\pm 0.4$ &$   2.8\pm 0.2$\\
Si III   &1301.149 &4.59 &$   2.8\pm 0.5$ &$   1.6\pm 0.2$ &     ...       \\
O I      &1302.169 &4.21 &$ 123.0\pm 1.3$ &$  89.9\pm 1.4$ &$  37.1\pm 0.6$\\
Si III   &1303.323 &4.59 &$   6.2\pm 0.5$ &$   2.7\pm 0.3$ &     ...       \\
Si II    &1304.370 &4.34 &$  13.3\pm 0.8$ &$   8.6\pm 0.4$ &$   2.6\pm 0.2$\\
O I      &1304.858 &4.21 &$ 192.8\pm 2.0$ &$  96.4\pm 1.2$ &$  40.8\pm 0.6$\\
O I      &1306.029 &4.21 &$ 192.3\pm 1.9$ &$  99.6\pm 1.2$ &$  37.9\pm 0.5$\\
Si II    &1309.276 &4.34 &$  22.2\pm 1.0$ &$  13.8\pm 0.4$ &$   4.5\pm 0.2$\\
C I      &1311.363 &4.25 &$   6.3\pm 0.4$ &$   3.4\pm 0.2$ &$   0.8\pm 0.2$\\
C II     &1334.532 &4.51 &$ 188.0\pm 6.0$ &$ 158.2\pm 1.4$ &$  50.6\pm 0.6$\\
C IIbl   &1335.68 &4.51 &$ 484.0\pm 10.0$ &$ 266.3\pm 1.9$ &$  80.9\pm 1.0$\\
O I      &1355.598 &4.20 &$  18.9\pm 0.6$ &$   9.9\pm 0.4$ &$   2.1\pm 0.2$\\
C I      &1355.844 &4.23 &$   7.0\pm 0.4$ &$   4.0\pm 0.2$ &$   0.5\pm 0.1$\\
O I      &1358.512 &4.20 &$   5.4\pm 0.3$ &$   3.0\pm 0.2$ &     ...       \\
C I      &1359.275 &4.23 &$   3.0\pm 0.3$ &$   1.5\pm 0.2$ &     ...       \\
Fe II    &1360.170 &4.29 &$   2.4\pm 0.3$ &$   1.4\pm 0.2$ &     ...       \\
Fe II    &1361.373 &4.29 &$   3.4\pm 0.3$ &$   1.9\pm 0.2$ &     ...       \\
C I      &1364.164 &4.23 &$   4.4\pm 0.4$ &$   2.2\pm 0.3$ &     ...       \\
O V      &1371.296 &5.33 &$  10.5\pm 0.5$ &$   3.8\pm 0.3$ &     ...       \\
Si IV    &1393.760 &4.75 &$ 230.0\pm 6.0$ &$ 136.1\pm 1.4$ &     ...       \\
O IV     &1401.157 &5.14 &$  18.1\pm 0.5$ &$   7.8\pm 0.5$ &     ...       \\
Si IV    &1402.773 &4.75 &$ 121.0\pm 5.0$ &$  72.5\pm 1.3$ &     ...       \\
S IV     &1404.808 &4.94 &$   6.9\pm 0.5$ &$   2.7\pm 0.3$ &     ...       \\
N IV     &1486.496 &5.08 &$   4.5\pm 1.5$ &$   3.2\pm 0.5$ &     ...       \\
Si II    &1526.707 &4.30 &$  50.8\pm 0.5$ &$  34.2\pm 0.9$ &     ...       \\
Si II    &1533.431 &4.30 &$  60.7\pm 0.5$ &$  39.9\pm 1.3$ &     ...       \\
C IV     &1548.204 &5.01 &$ 559.0\pm 11.0$ &$ 260.1\pm 3.1$ &    ...       \\
C IV     &1550.781 &5.01 &$ 282.0\pm 7.0$ &$ 136.5\pm 2.3$ &     ...       \\
He IIbl  &1640.44 &4.66 &$ 677.4\pm 6.4$ &$ 214.3\pm 3.3$ &      ...       \\
C Ibl    &1657    &4.16 &$ 524.5\pm 6.2$ &$ 261.5\pm 4.1$ &      ...       \\
O III    &1666.150 &4.90 &$  12.1\pm 1.5$ &$   7.0\pm 0.8$ &     ...       \\
\enddata
\end{deluxetable}

\clearpage

\begin{figure}
\plotfiddle{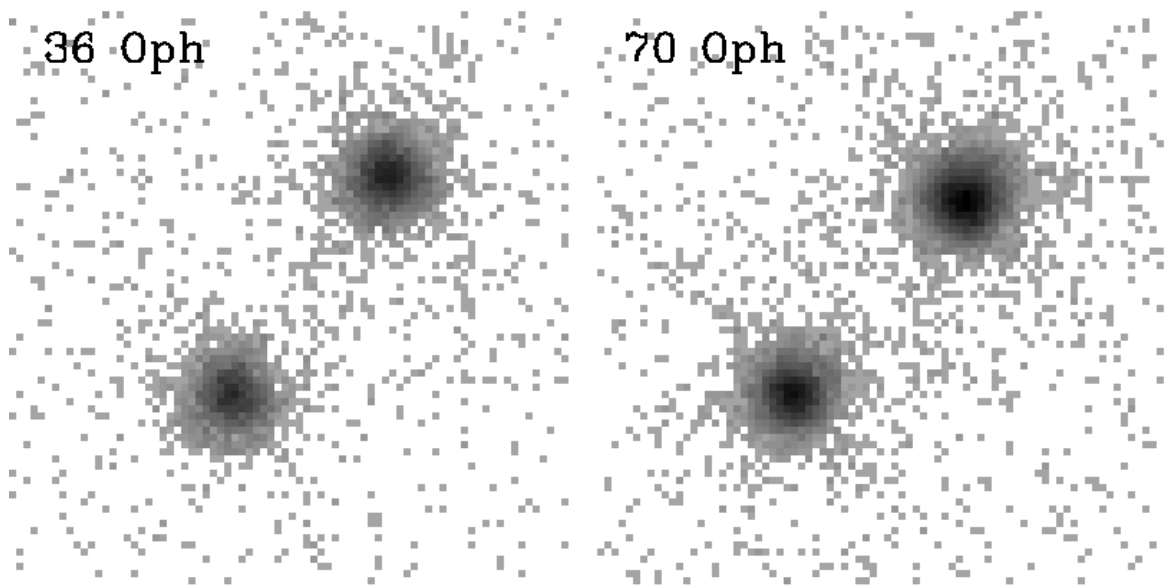}{3.5in}{0}{75}{75}{-230}{-290}
\caption{Zeroth-order images of 36~Oph and 70~Oph from {\em Chandra}
  LETG/HRC-S observations.  North is up in the figures and in each case
  the A component is the upper right constituent of the binary.  The
  scale is about $0.14^{\prime\prime}$ per pixel.}
\end{figure}

\clearpage

\begin{figure}
\plotfiddle{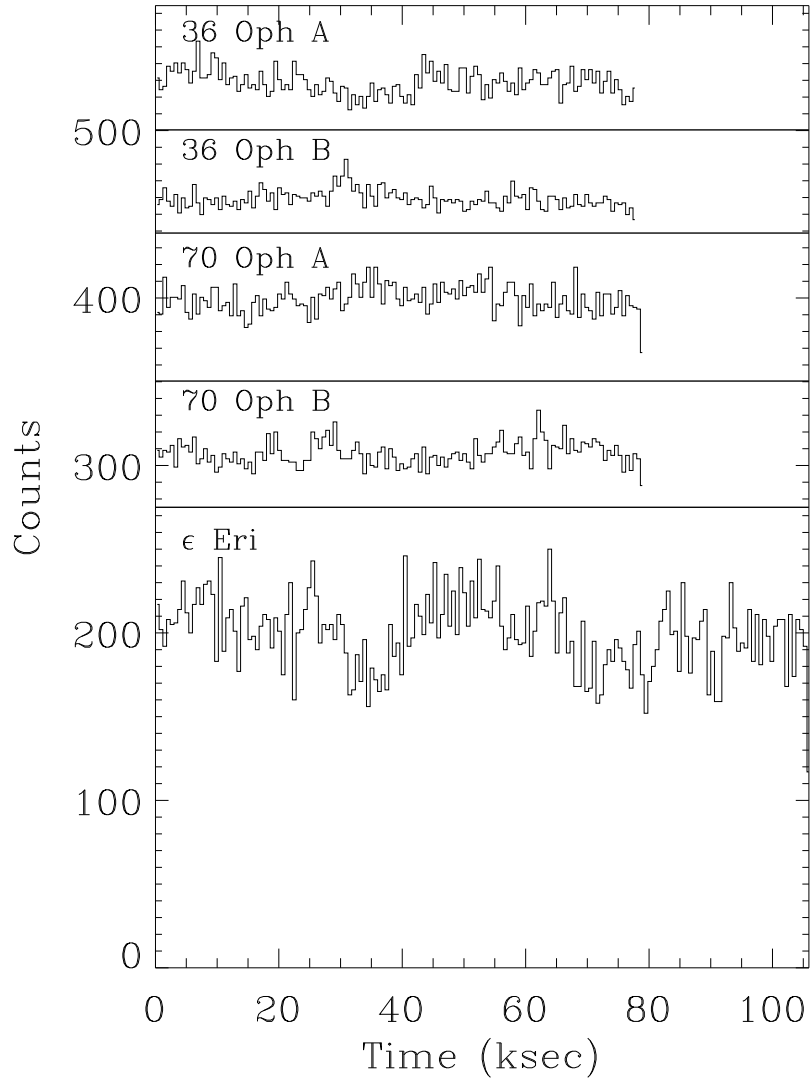}{3.5in}{90}{90}{90}{340}{-30}
\caption{X-ray light curves for our targets computed from the zeroth-order
  images in the {\em Chandra} LETG/HRC-S data, using 10 minute time bins.}
\end{figure}

\clearpage

\begin{figure}
\plotfiddle{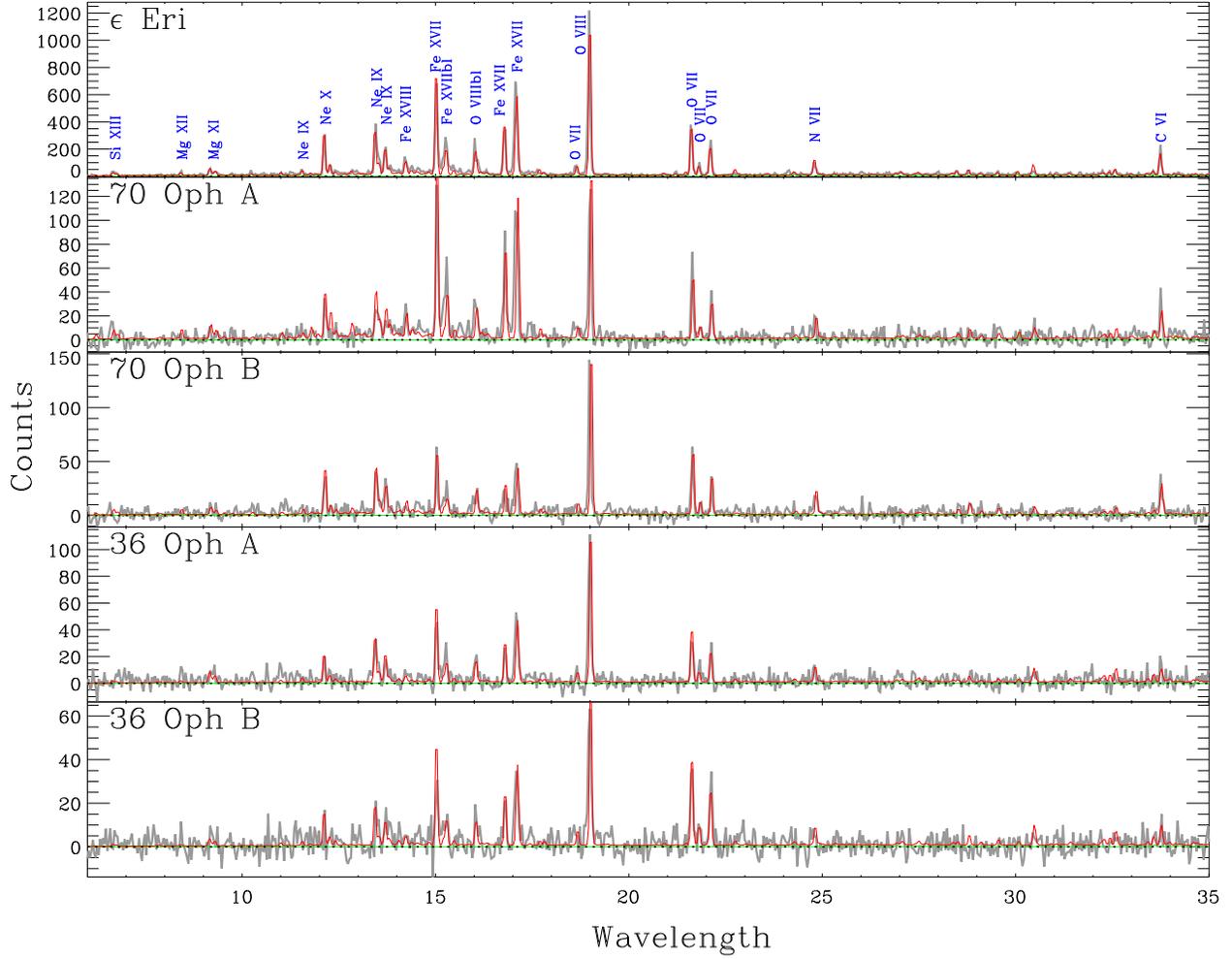}{3.5in}{90}{75}{75}{295}{0}
\caption{{\em Chandra} LETG/HRC-S spectra of 5 moderately active K dwarfs,
  with line identifications.  The spectra have been rebinned by a factor of
  3 to improve S/N.  For $\lambda > 35$~\AA, the spectra are
  also smoothed for the sake of appearance.  Red lines are
  synthetic spectra produced from derived emission measure distributions
  and coronal abundances (see \S4.4), and green lines indicate the
  contributions of higher orders ($2-5$) to the model spectra.}
\end{figure}

\clearpage

\setcounter{figure}{2}
\begin{figure}
\plotfiddle{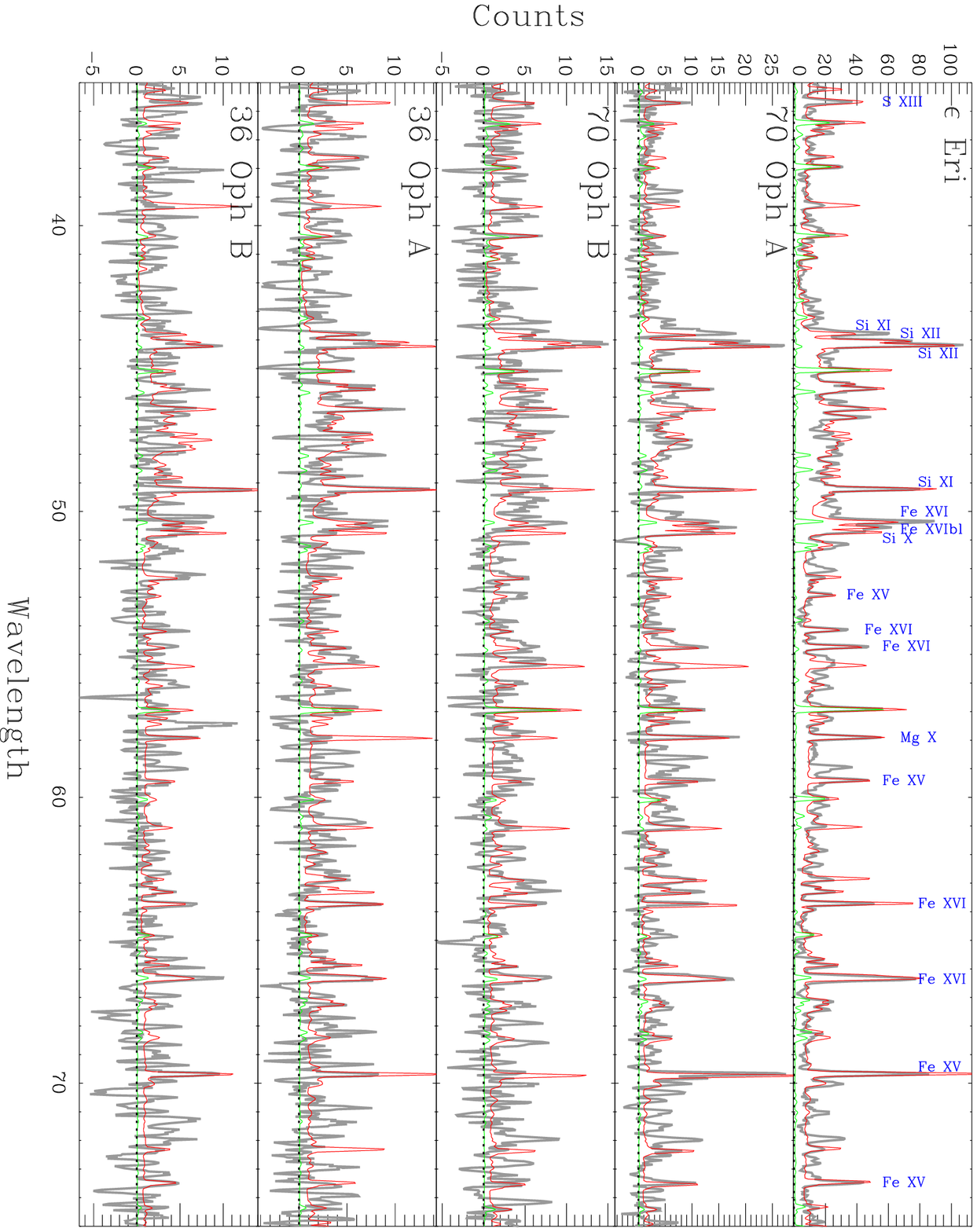}{3.5in}{90}{75}{75}{295}{0}
\caption{(continued)}
\end{figure}

\clearpage

\setcounter{figure}{2}
\begin{figure}
\plotfiddle{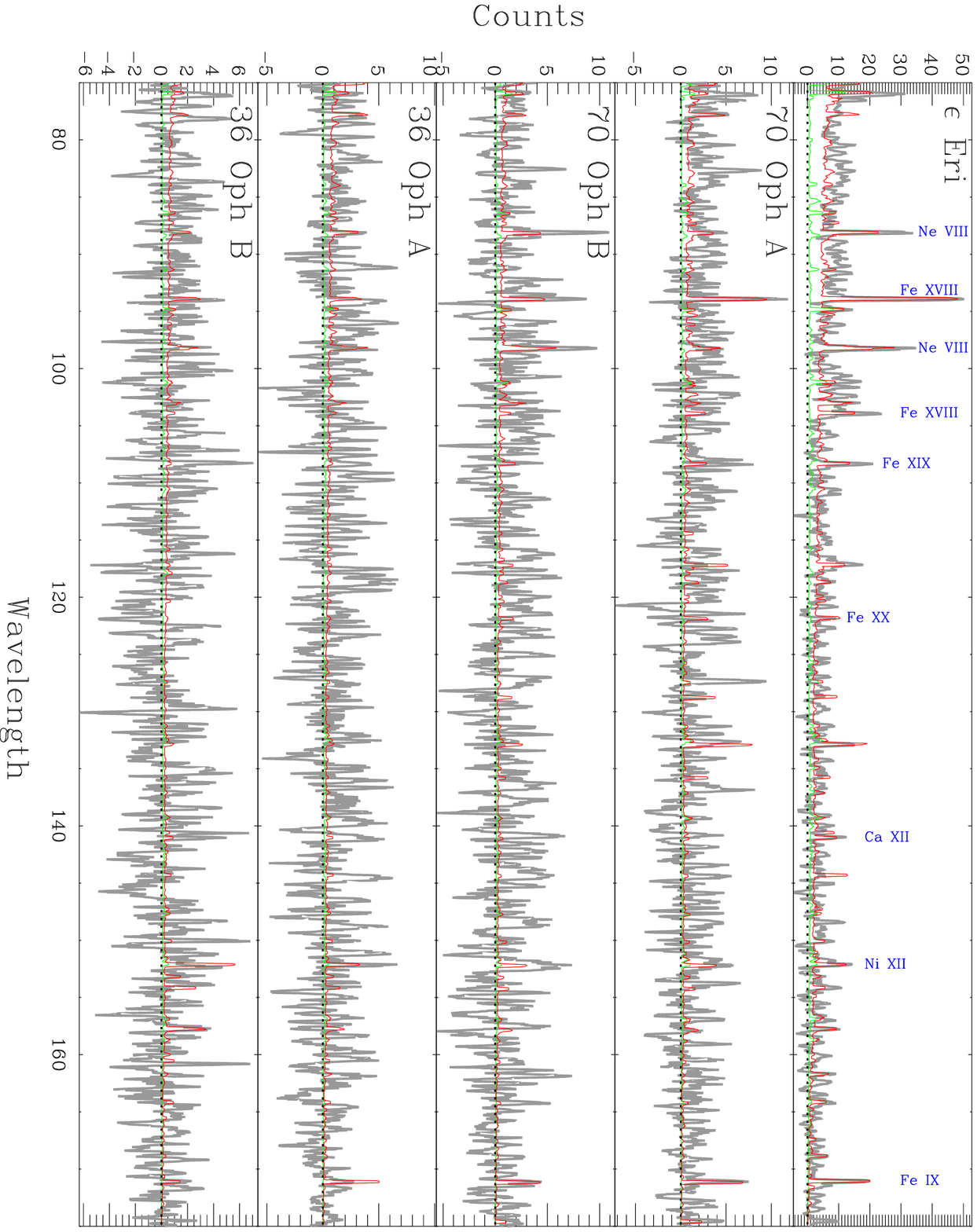}{3.5in}{90}{75}{75}{295}{0}
\caption{(continued)}
\end{figure}

\clearpage

\begin{figure}
\plotfiddle{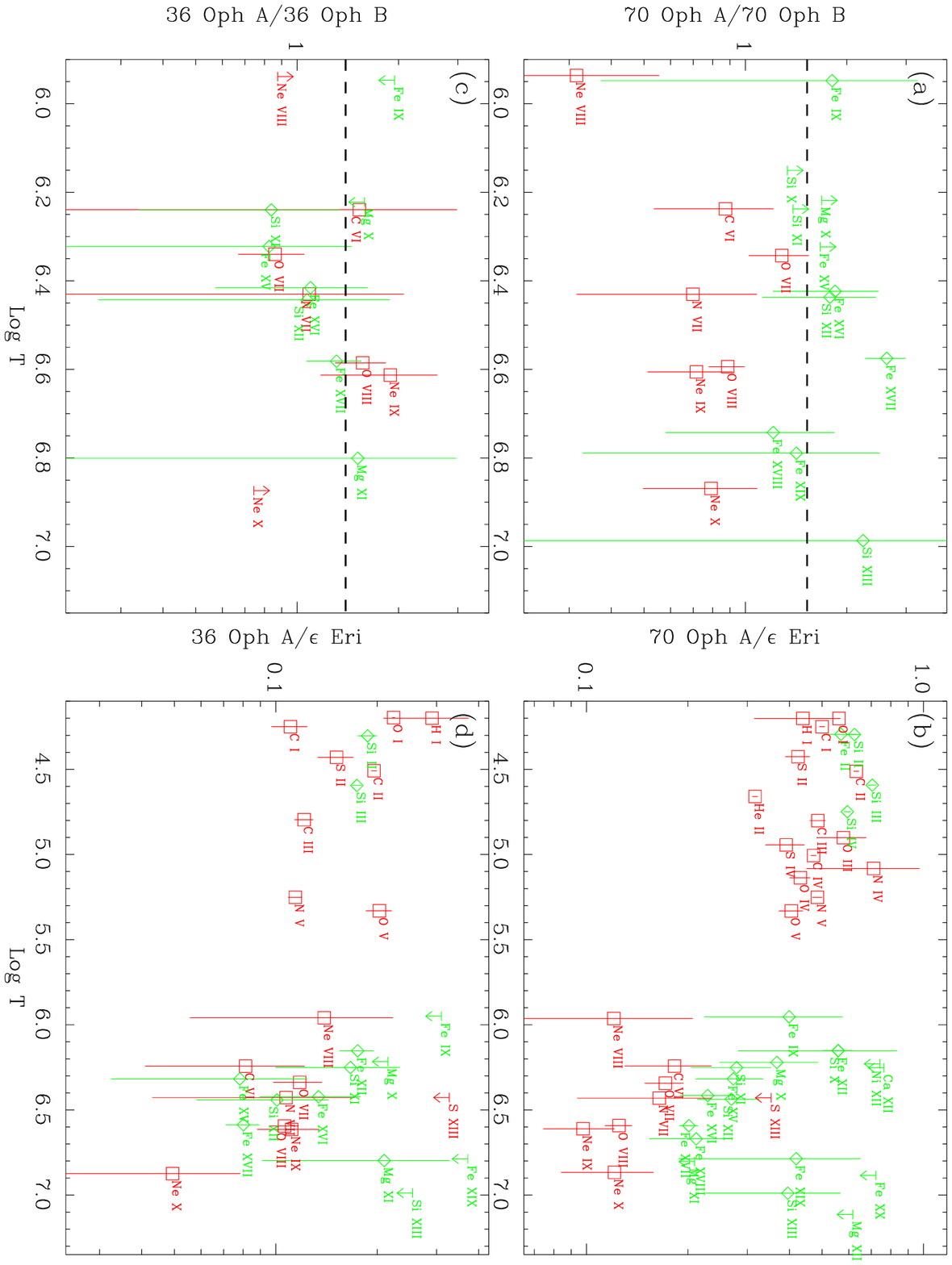}{3.5in}{90}{75}{75}{285}{-5}
\caption{Comparisons of emission line fluxes for various pairs of stars,
  illustrated by plotting flux ratios as a function of
  line formation temperature, with red and green data points indicating
  high-FIP and low-FIP elements, respectively.  Dashed lines in panels
  (a) and (c) indicate the flux ratios of zeroth-order {\em Chandra}
  LETG/HRC-S images.}
\end{figure}

\clearpage

\begin{figure}
\plotfiddle{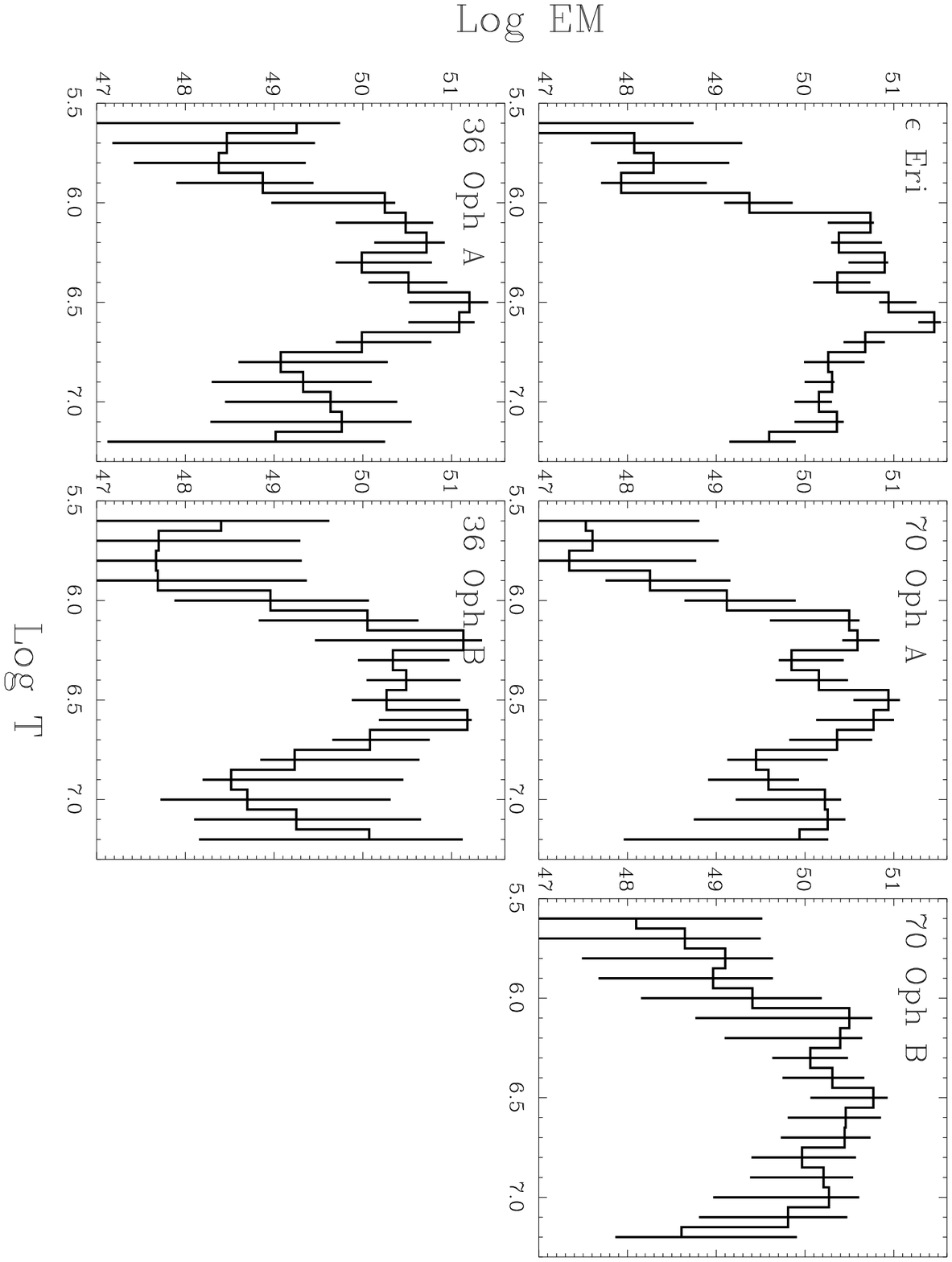}{3.5in}{90}{75}{75}{295}{0}
\caption{Emission measure distributions derived from {\em Chandra}
  LETG/HRC-S spectra, with EM in units of cm$^{-3}$.  Error bars are
  90\% confidence intervals.}
\end{figure}

\clearpage

\begin{figure}
\plotfiddle{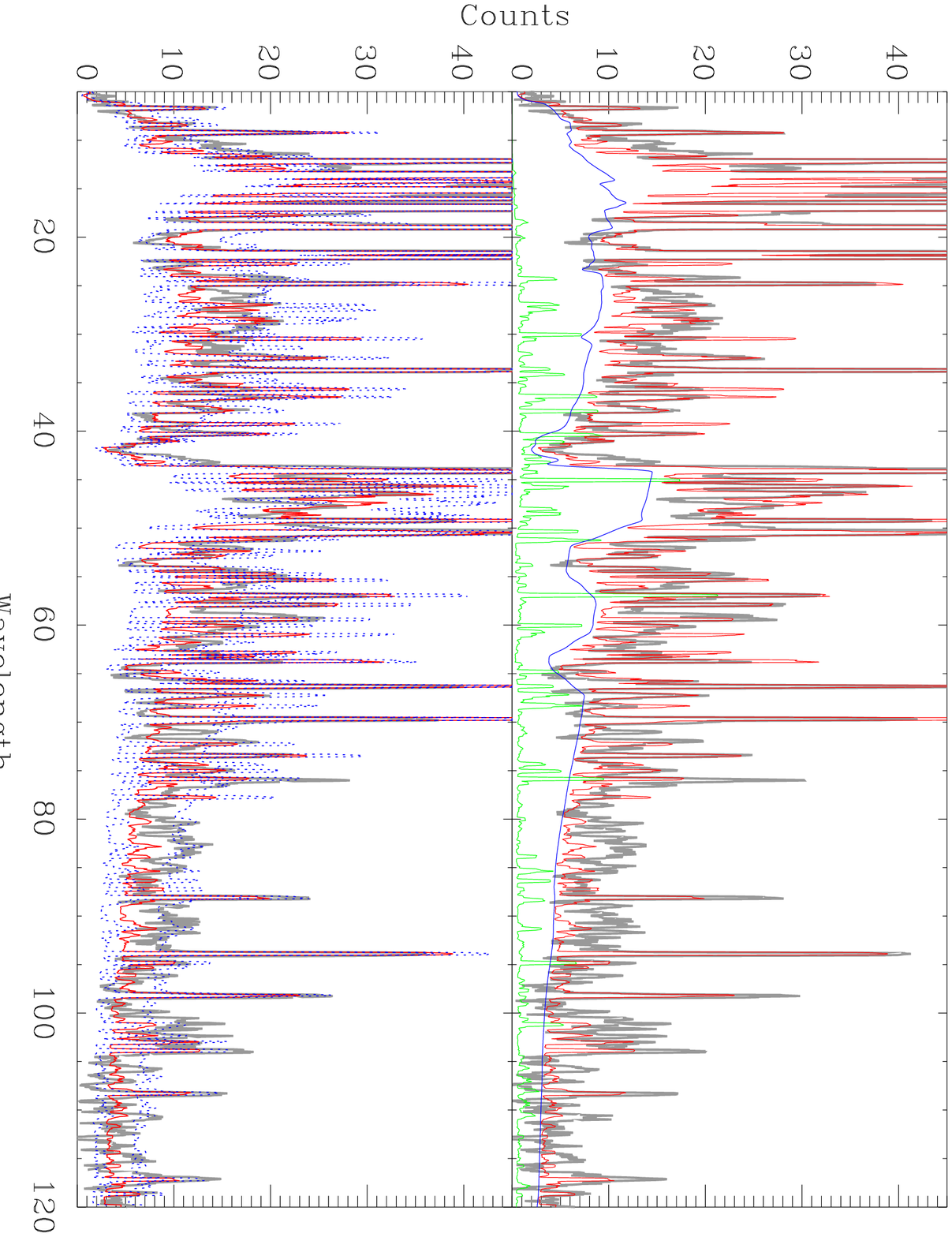}{3.5in}{90}{75}{75}{295}{0}
\caption{An illustration of how the line-to-continuum
  ratio is used to derive the absolute coronal Fe abundance, using
  $\epsilon$~Eri as an example.  The bottom panel compares the
  highly smoothed stellar spectrum with a synthetic spectrum (red line)
  computed from the emission measure distribution in
  Fig.~5, assuming a best-fit absolute Fe abundance
  of ${\rm [Fe/H]}=0.7{\rm [Fe/H]}_{\odot}$.  Dotted blue lines show the
  effect of raising or lowering this value by a factor of 2, where the
  higher line corresponds to lower [Fe/H].  The upper panel shows the
  same fit, and also shows explicitly the continuum (blue line) and higher
  order (green line) contributions to the total line-plus-continuum
  model spectrum (red line).}
\end{figure}

\clearpage

\begin{figure}
\plotfiddle{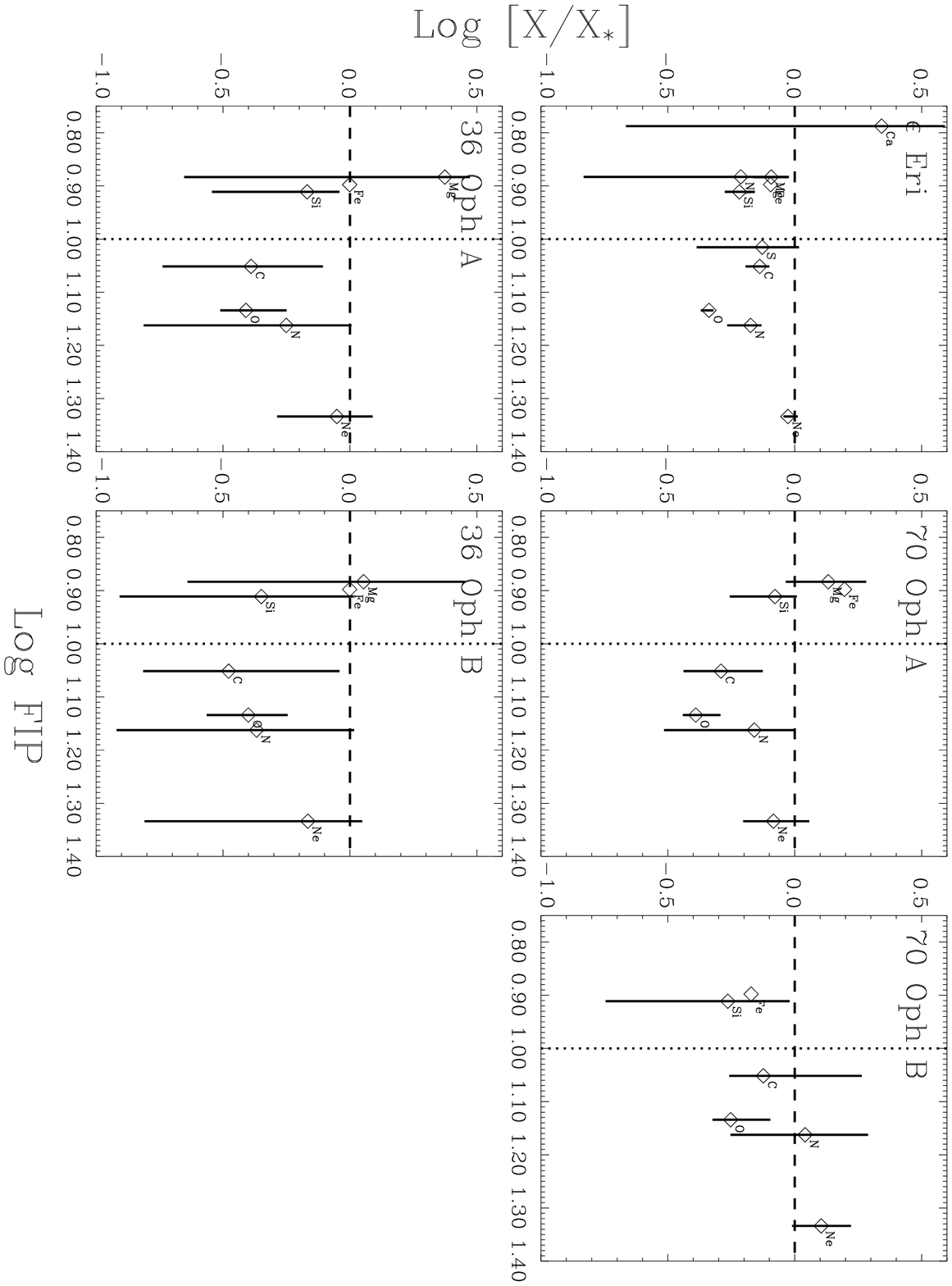}{3.5in}{90}{75}{75}{295}{0}
\caption{Coronal abundances relative to stellar photospheric abundances,
  where the error bars are 90\% confidence intervals.  The abundances
  are plotted versus FIP (in eV).  The dotted line crudely separates
  low-FIP from high-FIP elements.  Error bars are meant to indicate the
  uncertainties in the abundances {\em relative} to Fe, explaining why
  the Fe uncertainties are zero.}
\end{figure}

\clearpage

\begin{figure}
\plotfiddle{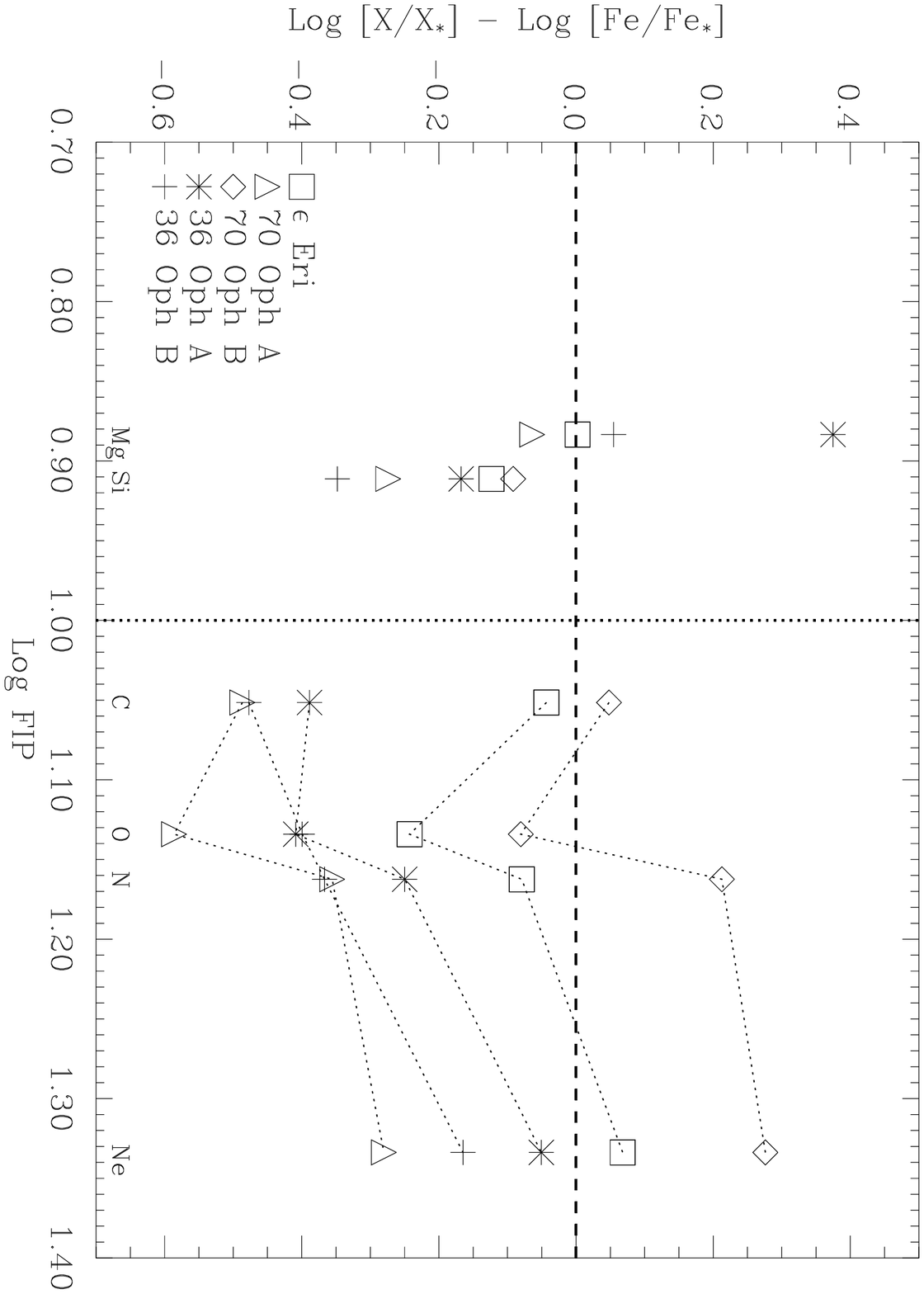}{3.5in}{90}{75}{75}{295}{0}
\caption{Coronal abundances from Fig.~7 divided by the Fe abundance,
  plotted versus FIP.  Dotted lines connect the high-FIP abundances for
  each star, suggesting the following sequence for increasing FIP effect:
  70~Oph~B, $\epsilon$~Eri, 36~Oph~A, 36~Oph~B, 70~Oph~A.}
\end{figure}


\clearpage

\begin{thebibliography}{}

\bibitem[Allende Prieto et al.(2004)]{cap04}
Allende Prieto, C., Barklem, P. S., Lambert, D. L., \& Cunha, K. 2004, A\&A,
  420, 183
\bibitem[Allende Prieto et al.(2001)]{cap01}
Allende Prieto, C., Lambert, D. L., \& Asplund, M. 2001, ApJ, 556, L63
\bibitem[Allende Prieto et al.(2002)]{cap02}
Allende Prieto, C., Lambert, D. L., \& Asplund, M. 2002, ApJ, 573, L137
\bibitem[Antia \& Basu(2005)]{hma05}
Antia, H. M., \& Basu, S. 2005, ApJ, 620, L129
\bibitem[Arnaud \& Rothenflug(1985)]{ma85}
Arnaud, M., \& Rothenflug, R. 1985, A\&AS, 60, 425
\bibitem[Asplund et al.(2005)]{ma05}
Asplund, M., Grevesse, N., \& Sauval, A. J. 2005, in Cosmic Abundances as
  Records of Stellar Evolution and Nucleosynthesis, ed. T. G. Barnes III \&
  F. N. Bash (San Francisco:  ASP), 25
\bibitem[Audard et al.(2001)]{ma01}
Audard, M., Behar, E., G\"{u}del, M., Raassen, A. J. J., Porquet, D.,
  Mewe, R., Foley, C. R., \& Bromage, G. E. 2001, A\&A, 365, L329
\bibitem[Audard et al.(2003)]{ma03}
Audard, M., G\"{u}del, M., Sres, A., Raassen, A. J. J., \& Mewe, R. 2003,
  A\&A, 398, 1137
\bibitem[Ayres et al.(2003)]{tra03}
Ayres, T. R., Brown, A.; Harper, G. M., Osten, R. A., Linsky, J. L., Wood,
  B. E., \& Redfield, S. 2003, ApJ, 583, 963
\bibitem[Bahcall et al.(2005a)]{jnb05a}
Bahcall, J. N., Basu, S., Pinsonneault, M., \& Serenelli, A. M. 2005a, ApJ,
  618, 1049
\bibitem[Bahcall et al.(2005b)]{jnb05b}
Bahcall, J. N., Basu, S., \& Serenelli, A. M. 2005b, ApJ, 631, 1281
\bibitem[Ball et al.(2005)]{bb05}
Ball, B., Drake, J. J., Lin, L., Kashyap, V., Laming, J. M., \&
  Garc\'{i}a-Alvarez, D. 2005, ApJ, 634, 1336
\bibitem[Barnes et al.(1978)]{tgb78}
Barnes, T. G., Evans, D. S., \& Moffett, T. J. 1978, MNRAS, 183, 285
\bibitem[Basu \& Antia(2004)]{sb04}
Basu, S., \& Antia, H. M. 2004, ApJ, 606, L85
\bibitem[Brinkman et al.(2001)]{acb01}
Brinkman, A. C., et al. 2001, A\&A, 365, L324
\bibitem[Dere et al.(1997)]{kpd97}
Dere, K. P., Landi, E., Mason, H. E., Monsignori Fossi, B. C., \& Young,
  P. R. 1997, A\&AS, 125, 149
\bibitem[Donahue et al.(1996)]{rad96}
Donahue, R. A., Saar, S. H., \& Baliunas, S. L. 1996, ApJ, 466, 384
\bibitem[Drake et al.(2001)]{jjd01}
Drake, J. J., Brickhouse, N. S., Kashyap, V., Laming, J. M.,
  Huenemoerder, D. P., Smith, R., \& Wargelin, B. J. 2001, ApJ, 548, L81
\bibitem[Drake et al.(1995)]{jjd95}
Drake, J. J., Laming, J. M., \& Widing, K. G. 1995, ApJ, 443, 393
\bibitem[Drake et al.(1997)]{jjd97}
Drake, J. J., Laming, J. M., \& Widing, K. G. 1997, ApJ, 478, 403
\bibitem[Drake \& Testa(2005)]{jjd05}
Drake, J. J., \& Testa, P. 2005, Nature, 436, 525
\bibitem[Feldman \& Laming(1994)]{uf94}
Feldman, U., \& Laming, J. M. 1994, ApJ, 434, 370
\bibitem[Feldman \& Laming(2000)]{uf00}
Feldman, U., \& Laming, J. M. 2000, Phys. Scr., 61, 222
\bibitem[Fludra \& Schmelz(1995)]{af95}
Fludra, A., \& Schmelz, J. T. 1995, ApJ, 447, 936
\bibitem[Gloeckler \& Geiss(2004)]{gg04}
Gloeckler, G., \& Geiss, J. 2004, Adv.~Space~Res., 34, 53
\bibitem[Grevesse \& Sauval(1998)]{ng98}
Grevesse, N., \& Sauval, A. J. 1998, Space Sci. Rev., 85, 161
\bibitem[G\"{u}del(2004)]{mg04}
G\"{u}del, M. 2004, Astron. Astrophys. Rev., 12, 71
\bibitem[G\"{u}del et al.(2001)]{mg01}
G\"{u}del, M., et al. 2001, A\&A, 365, L336
\bibitem[Huenemoerder et al.(2001)]{dph01}
Huenemoerder, D. P., Canizares, C. R., \& Schulz, N. S. 2001, ApJ, 559, 1135
\bibitem[Huenemoerder et al.(2003)]{dph03}
Huenemoerder, D. P., Canizares, C. R., Schulz, N. S., Drake, J. J., \&
  Sanz-Forcada, J. 2003, ApJ, 595, 1131
\bibitem[Kashyap \& Drake(1998)]{vk98}
Kashyap, V. \& Drake, J. J. 1998, ApJ, 503, 450
\bibitem[Kashyap \& Drake(2000)]{vk00}
Kashyap, V. \& Drake, J. J. 2000, BASI, 28, 475
\bibitem[Laming(2004)]{jml04}
Laming, J. M. 2004, ApJ, 614, 1063
\bibitem[Laming \& Drake(1999)]{jml99}
Laming, J. M., \& Drake, J. J. 1999, ApJ, 516, 324
\bibitem[Laming et al.(1995)]{jml95}
Laming, J. M., Drake, J. J., \& Widing, K. G. 1995, ApJ, 443, 416
\bibitem[Laming et al.(1996)]{jml96}
Laming, J. M., Drake, J. J., \& Widing, K. G. 1996, ApJ, 462, 948
\bibitem[Mazzotta et al.(1998)]{pm98}
Mazzotta, P., Mazzitelli, G., Colafrancesco, S., \& Vittorio, N. 1998, A\&AS,
  133, 403
\bibitem[McKenzie(1987)]{dlm87}
McKenzie, D. L. 1987, ApJ, 322, 512
\bibitem[Morrison \& McCammon(1983)]{rm83}
Morrison, R., \& McCammon, D. 1983, ApJ, 270, 119
\bibitem[Ness et al.(2002)]{jun02}
Ness, J. -U., Schmitt, J. H. M. M., Burwitz, V., Mewe, R., Raassen, A. J. J.,
  van der Meer, R. L. J., Predehl, P., \& Brinkman, A. C. 2002, A\&A,
  394, 911
\bibitem[Noyes et al.(1984)]{rwn84}
Noyes, R. W., Hartmann, L. W., Baliunas, S. L., Duncan, D. K., \& Vaughan,
  A. H. 1984, ApJ, 279, 763
\bibitem[Perryman et al.(1997)]{macp97}
Perryman, M. A. C., et al. 1997, A\&A, 323, L49
\bibitem[Porquet et al.(2001)]{dp01}
Porquet, D., Mewe, R., Dubau, J., Raassen, A. J. J., \& Kaastra, J. S. 2001,
  A\&A, 376, 1113
\bibitem[Raassen et al.(2002)]{ar02}
Raassen, A. J. J., et al. 2002, A\&A, 389, 228
\bibitem[Raassen et al.(2003)]{ar03}
Raassen, A. J. J., Ness, J. -U., Mewe, R., van der Meer, R. L. J.,
  Burwitz, V., \& Kaastra, J. S. 2003, A\&A, 400, 671
\bibitem[Sanz-Forcada et al.(2004)]{jsf04}
Sanz-Forcada, J., Favata, F., \& Micela, G. 2004, A\&A, 416, 281
\bibitem[Sanz-Forcada et al.(2003)]{jsf03}
Sanz-Forcada, J., Maggio, A., \& Micela, G. 2003, A\&A, 408, 1087
\bibitem[Schmelz et al.(2005)]{jts05}
Schmelz, J. T., Nasraoui, K., Roames, J. K., Lippner, L. A., \& Garst, J. W.
  2005, ApJ, 634, L197
\bibitem[Schmelz et al.(1996)]{jts96}
Schmelz, J. T., Saba, J. L. R., Ghosh, D., \& Strong, K. T. 1996, ApJ,
  473, 519
\bibitem[Schmitt \& Liefke(2004)]{js04}
Schmitt, J. H. M. M., \& Liefke, C. 2004, A\&A, 417, 651
\bibitem[Shemi(1991)]{as91}
Shemi, A. 1991, MNRAS, 251, 221
\bibitem[Sim \& Jordan(2005)]{sas05}
Sim, S. A., \& Jordan, C. 2005, MNRAS, 361, 1102
\bibitem[Telleschi et al.(2005)]{at05}
Telleschi, A., G\"{u}del, M., Briggs, K., Audard, M., Ness, J. -U., \&
  Skinner, S. L. 2005, ApJ, 622, 653
\bibitem[Veck \& Parkinson(1991)]{njv81}
Veck, N. J., \& Parkinson, J. H. 1981, MNRAS, 197, 41
\bibitem[von Steiger \& Geiss(1989)]{rvs89}
von Steiger, R., \& Geiss, J. 1989, A\&A, 225, 222
\bibitem[von Steiger et al.(2000)]{rvs00}
von Steiger, R., et al. 2000, J.~Geophys.~Res., 105, 27217
\bibitem[Widing \& Feldman(2001)]{kgw01}
Widing, K. G., \& Feldman, U. 2001, ApJ, 555, 426
\bibitem[Wood et al.(2005a)]{bew05a}
Wood, B. E., M\"{u}ller, H. -R., Zank, G. P., Linsky, J. L., \&
  Redfield, S. 2005a, ApJ, 628, L143
\bibitem[Wood et al.(2005b)]{bew05b}
Wood, B. E., Redfield, S., Linsky, J. L., M\"{u}ller, H. -R., \& Zank, G. P.
  2005b, ApJS, 159, 118
\bibitem[Young(2005)]{pry05}
Young, P. R. 2005, A\&A, 444, L45
\bibitem[Young et al.(2003)]{pry03}
Young, P. R., Del Zanna, G., Landi, E., Dere, K. P., Mason, H. E., \&
  Landini, M. 2003, ApJS, 144, 135

\end{thebibliography}
\end{document}